\documentclass[useAMS,usenatbib]{mn2e}
\usepackage{graphicx, txfonts}
\usepackage{url}
\usepackage{multirow}

\usepackage[usenames,dvipsnames]{color}
\usepackage[normalem]{ulem}

\newcommand{\CoI}{\textrm{Co}\,\textsc{i}}
\newcommand{\CoII}{\textrm{Co}\,\textsc{ii}}
\newcommand{\CrI}{\textrm{Cr}\,\textsc{i}}
\newcommand{\CrII}{\textrm{Cr}\,\textsc{ii}}

\newcommand{\FeII}{Fe\,\textsc{ii}}

\newcommand{\HI}{\textrm{H}\,\textsc{i}}

\newcommand{\Lya}{Ly$\alpha$}

\newcommand{\NHI}{$N(\textrm{H}\,\textsc{i})$}

\newcommand{\NiII}{Ni\,\textsc{ii}}

\newcommand{\SII}{S\,\textsc{ii}}
\newcommand{\SiII}{Si\,\textsc{ii}}

\newcommand{\TiI}{Ti\,\textsc{i}}
\newcommand{\TiII}{Ti\,\textsc{ii}}
\newcommand{\ZnII}{\textrm{Zn}\,\textsc{ii}}

\title[The Fe-peak elements in low metallicity DLAs]
{The explosion energy of early stellar populations: The Fe-peak element ratios in low metallicity damped Ly$\alpha$ systems\thanks{
Based on observations collected at the
European Organisation for Astronomical Research 
in the Southern Hemisphere, Chile [Proposals 67.A-0078(A),
69.A-0613(A), 083.A-0042(A), and 085.A-0109(A)], and 
at the W.M. Keck Observatory
which is operated as a scientific partnership among the California Institute of 
Technology, the University of California and the National Aeronautics and Space 
Administration. The Observatory was made possible by the generous financial
support of the W.M. Keck Foundation.
Keck telescope time was partially granted by NOAO,
through the Telescope System Instrumentation Program
(TSIP). TSIP is funded by NSF.
}}

\author[Cooke et al.]{Ryan Cooke,$^{1,2}$\thanks{Morrison Fellow, email: rcooke@ucolick.org}
Max Pettini,$^{2}$
Regina A. Jorgenson,$^3$
Michael T. Murphy,$^{4}$
\newauthor Gwen C. Rudie$^5$ and
Charles C. Steidel$^5$
\\
$^1$Department of Astronomy and Astrophysics, University of California,
Santa Cruz, CA 95064, USA\\
$^2$Institute of Astronomy, Madingley Road, Cambridge, CB3 0HA\\
$^3$Institute for Astronomy, University of Hawaii, 2680 Woodlawn Drive, Honolulu, HI 96822, USA\\
$^4$Centre for Astrophysics and Supercomputing, Swinburne
University of Technology, Hawthorn, Victoria 3122, Australia\\
$^5$California Institute of Technology, MS 249-17, Pasadena, CA 91125, USA\\
\\
}

\begin{document}

\date{Accepted . Received ; in original form }
\pagerange{\pageref{firstpage}--\pageref{lastpage}} 
\pubyear{2011}

\maketitle

\label{firstpage}

\begin{abstract}
The relative abundances of the Fe-peak elements (Ti--Zn) at the lowest metallicities
are intimately linked to the physics of core-collapse supernova explosions. With
a sample of 25 very metal-poor damped \Lya\ systems, we investigate the trends
of the Fe-peak element ratios with metallicity. 
For nine of the 25 DLAs,
a direct measurement (or useful upper limit) of 
one or more of the Ti,Cr,Co,Ni,Zn/Fe abundance ratios 
could be determined from detected absorption lines. 
For the remaining systems (without detections),
we devised a new form of spectral stacking to estimate 
the typical Fe-peak element ratios of
the DLA population in this metallicity regime.
We compare these data to analogous measurements in metal-poor stars
of the Galactic halo and to detailed calculations of explosive
nucleosynthesis in metal-free stars.
We conclude that most of the DLAs in our sample 
were enriched by stars that released an energy
of $\lesssim1.2\times10^{51}$ erg when they exploded as
core-collapse supernovae. 
Finally, we discuss the exciting prospect of measuring Fe-peak
element ratios in damped \Lya\ systems with Fe/H\,$< 1/1000$ of solar 
when 30-m class telescopes become available. Only then will we be able
to pin down the energy that was released by the supernovae of the first stars.
\end{abstract}

\begin{keywords}
galaxies: abundances $-$ galaxies: evolution $-$
quasars: absorption lines $-$ supernovae: general
\end{keywords}

%%%%%%%%%%%%%%%%%%%%%%
\section{Introduction}
%%%%%%%%%%%%%%%%%%%%%%

The physical mechanism that drives a
core-collapse supernova (CCSN) explosion
is a challenging mystery that remains unsolved. However,
it has been widely recognised for some time
that the relative abundances of the Fe-peak elements
(Ti--Zn),
which predominantly originate from explosive
nucleosynthesis (see e.g. \citealt{WooWea95}), can
provide unique insights into the inner workings of a
massive star in its death throes \citep{UmeNom02}.
In particular, the
ratios of the Fe-peak element yields depend sensitively
on the progenitor mass, the explosion energy, and the
amount of material that falls back onto the remnant
(i.e. the mass-cut).
The uncertain physics of the explosion mechanism 
forces model nucleosynthesis
calculations to parameterise the explosion.
There are two favoured prescriptions to define
the central engine of the explosion: the first mechanism
involves depositing momentum at a given mass coordinate
(\citealt{WooWea95,ChiLim04,HegWoo10}), whereas
the second is a deposition of thermal energy \citep{Nak01}.
Both prescriptions appear to give consistent yields for
the lighter $\alpha$-elements,
although they produce quite different abundances of the Fe-peak
elements \citep{AudBarThi01}. This is not surprising, given
that the Fe-peak elements are intimately linked to the details
of the explosion.

It is reassuring that one of these prescriptions (momentum deposition)
is able to do a remarkably good job at reproducing the large variety of
abundance patterns that massive Population III stars (10--100\,M$_{\odot}$)
encoded in the second generation of stars (see, for example, the comparisons
performed by \citealt{HegWoo10}). These early stellar populations have
largely been discovered through the dedicated HK \citep{BeePreShe92}
and Hamburg/European Southern Observatory \citep[HES;][]{Chr01} surveys, with detailed abundance analyses
being notably conducted by \citet{McW95}, \citet{Cay04}, and \citet{Lai08}
and references therein.
The observations have uncovered some unexpected trends in the
relative abundance of Fe-peak elements with decreasing metallicity,
which we now briefly discuss.

The first detailed study of the Fe-peak elements in
stars with metallicities $<1/100$ of solar ([Fe/H]\,$< -2$) was
conducted by \citet{McW95}. Their sample comprised
the 33 most metal-poor halo stars known at the time,
extending down to  [Fe/H]\,$\sim-4.0$.
These authors
noted changes in the general trends of some of the Fe-peak
element ratios for metallicities  [Fe/H]\,$\lesssim -2.4$. 
In particular, they found that the Cr/Fe ratio,
which is roughly solar in stars with
[Fe/H]\,$\gtrsim -2.4$, decreases to [Cr/Fe]\,$\simeq -0.5$ at
[Fe/H]\,$=-4.0$, albeit with some scatter. 
This trend was later confirmed by 
\citet{Cay04} and \citet{Lai08}, whose data 
exhibited reduced scatter.
Unfortunately, there is still a worrisome
disagreement between the Cr/Fe ratios
deduced from either the \CrI\ or \CrII\ lines; 
differences of up to $0.3\,$dex 
for metallicities [Fe/H]\,$< -2.4$ may be expected
\citep{SobLawSne07}, and the true Cr/Fe 
relative abundance
in the low metallicity regime may in fact not be 
significantly different from the solar value.

The Ni/Fe ratio, on the other hand, exhibits no discernible
trend away from the solar value down to the lowest metallicities
probed by the above surveys. This is quite unlike the
Zn/Fe ratio, which is roughly solar in the 
metallicity interval $-2.0 <  {\rm  [Fe/H]} < 0.0$,
but then gradually increases  with decreasing metallicity 
reaching [Zn/Fe]\,$\simeq +0.7$ at [Fe/H]\,$= -4.0$
\citep{Pri00,Cay04}. 
If such a behaviour were due to CCSN nucleosynthesis,
the excess production of zinc is likely the result of a strong
$\alpha$-rich freeze out \citep{WooWea95}, which in turn is
linked to the explosion energy. In fact, \citet{UmeNom02}
invoked the dominance of hypernovae---a CCSN explosion
with an order-of-magnitude more energy released than normal---at the
lowest metallicities to explain the elevated [Zn/Fe] values.
This interpretation is also consistent with the observed rise
in the [Co/Fe] values exhibited by the same group of stars.

Despite these significant advances in the stellar
abundance measurements, it would be highly
advantageous to have complementary 
probes to independently confirm the trends uncovered
in the most metal-poor stars. This will ensure that
we are not biased by systematics that may affect the 
modelling of the absorption lines in the stellar atmospheres.
With this goal in mind, we and other groups have recently
focussed our attention on the most metal-poor clouds 
of neutral gas at redshifts $z\sim 2$--3
(\citealt{Pet08,Pen08,Coo11a,Coo11b}; see also \citealt{Bec12}). 
Such clouds are identified amongst
the damped \Lya\ systems (DLAs), absorbers
with neutral hydrogen column densities 
$N$(H\,{\sc i})\,$\geq 10^{20.3}$\,atoms~cm$^{-2}$
(\citealt{Wol86}; for a review of this topic see \citealt{WolGawPro05}).
At such large column densities the clouds are self-shielded
from the metagalactic ionizing background, and most elements 
are concentrated in a single dominant ionization state,
usually the neutrals or the first ions (e.g. \citealt{Vla01}).
This simple ionization structure makes the measurement of element
abundances straightforward.
In a recently completed survey of 22 DLAs with
[Fe/H] $< -2.0$, \citet{Coo11b} noted a good agreement 
in the abundances of C, O and Fe between halo
stars and DLAs of comparable metallicity. 
This is perhaps not surprising, since high redshift 
very metal-poor (VMP) DLAs 
likely contain the gas from which the most metal-poor stars 
may have later condensed. As discussed by \citet{PetCoo13},
VMP DLAs  are therefore an ideal, complementary
probe to measure the relative abundances of the Fe-peak elements,
and compare them to those found in Galactic halo stars.

This is the subject of the present paper, where
we measure Fe-peak element ratios in a slightly expanded
version of the sample of VMP DLAs considered by \citet{Coo11b}. 
This is a difficult prospect because most Fe-peak elements are
less abundant than Fe by one to two orders of magnitude and
the absorption lines of their dominant ion stages in VMP DLAs
are generally too weak to be detected with current 
observational facilities. 
To circumvent this issue, we introduce a form
of spectral stacking that allows us
to measure typical Fe-peak element ratios for the whole
sample of VMP DLAs. In Section~2, we briefly describe the
sample. In Section~\ref{sec:analysis} we introduce the spectral
stacking technique and present the results
of our analysis. We then compare in Section~\ref{sec:stars}
the DLA Fe-peak element
ratios with those measured in metal-poor halo stars.
Finally, in Section~\ref{sec:models} we compare these ratios
with theoretical yields calculated with the current generation 
of explosive nucleosynthesis models for metal-free stars,
before concluding  in Section~\ref{sec:conc}.

%%%%%%%%%%%%%%%%%%%%%%%%%%%%%%%%%%%%%%%%%
\section{The Sample}
\label{sec:sam}
%%%%%%%%%%%%%%%%%%%%%%%%%%%%%%%%%%%%%%%%%

%% Rejected Systems:
% J0003m2603: Measures reported by Molaro et al. (2000)
% J0311-1722 : No FeII lines
% J1016+4040 : No FeII lines
% J0307-4945 : Poor wavelength calibration

The majority of the VMP DLAs considered in the present work
were drawn from the survey by
\citet{Coo11b} which comprises 22 DLAs with metallicities
less than 1/100 of solar.
We excluded three DLAs from that sample. Two
of these were excluded because no \FeII\ absorption
lines were detected. 
In the third case, we found that 
the wavelength calibration
was too uncertain to reliably stack portions
of the spectrum for the purpose of detecting a signal from weak absorption
lines (see Section~\ref{sec:stacktech}).
We augmented the VMP DLA sample of \citet{Coo11b}
with two further systems from our own work,
one published \citep{Coo12}  and one which 
is the subject of a forthcoming study
(Cooke et al., in preparation). Furthermore,
we have also included 
five VMP DLAs
reported in the literature that didn't meet the selection cuts
of \citet{Coo11b}, but show clear detections of
absorption lines from  Fe-peak elements. 
Thus the total sample considered here
consists of  25 VMP DLAs. Details of these
systems are collected in Table~\ref{tab:targets}.

\begin{table}
    \caption{\textsc{The DLA sample}}
    \begin{tabular}{@{}lcccccc}
    \hline
    \hline
   \multirow{2}{*}{QSO Name}
& \multirow{2}{*}{$z_{\rm em}$}
& \multirow{2}{*}{$z_{\rm abs}$}
& \multirow{2}{*}{log \Large{$\frac{N{\rm (Fe\,\normalsize{\textsc{ii}})}}{{\rm cm}^{-2}}$}}
& \multirow{2}{*}{log \Large{$\frac{N{\rm (H\,\normalsize{\textsc{i}})}}{{\rm cm}^{-2}}$}}
& \multirow{2}{*}{S/N$^{\rm a}$}\\
&
&
&
&
&\\
    \hline
J0035$-$0918    &     2.420   &    2.340097     &     12.99     &       20.55   &   17 \\
Q0112$-$306     &     2.985   &   2.418440    &     13.33     &       20.50   &   23  \\
J0140$-$0839     &   3.716     &   3.696600     &     12.77     &       20.75   &  31  \\
J0831$+$3358    &   2.427     &    2.303642    &      13.33    &        20.25   &  15 \\
Q0913$+$072     &   2.785     &    2.618376    &      12.99    &        20.34   &  47   \\
J1001$+$0343    &   3.198     &    3.078413   &       12.50    &        20.21   &  40 \\
J1037$+$0139    &  3.059      &    2.704870    &      13.53    &        20.50   &  45 \\
Q1108$-$077     &   3.922     &    3.607670     &     13.88     &       20.37   &   49  \\
J1112$+$1333   &   2.420     &     2.270940    &      13.58    &        20.44   &  53  \\
J1337$+$3152   &   3.174     &     3.167667    &      13.14    &        20.41   &  19   \\
J1340$+$1106   &   2.914     &     2.507918    &      13.49    &        20.09   &  50 \\
J1340$+$1106   &   2.914     &    2.795826    &      14.32     &       21.00   &   50 \\
J1358$+$6522    &  3.173      &    3.067295   &       13.10    &        20.47   &  30    \\
J1419$+$0829   &   3.030     &     3.049840     &     13.54    &        20.391   & 43  \\
J1419$+$5923  &    2.321    &      2.247296     &     14.35    &        20.95   &   18   \\
J1558$-$0031   &    2.823    &     2.702389     &     14.11     &       20.67   &    31  \\
J1558$+$4053  &   2.635     &      2.553766    &      13.07    &        20.30   &   18  \\
J1944$+$7705  &   2.994     &      2.844300    &      13.24    &        20.27   &  110   \\
Q2059$-$360    &  3.090      &    3.082833     &     14.48     &       20.98   &    27  \\
J2155$+$1358  &  4.256      &     4.212210    &      12.93    &        19.61   &   22   \\
Q2206$-$199    &   2.559     &     2.076227     &     13.33      &      20.43   &   100   \\
    \hline
    \end{tabular}
    \smallskip

 $^{\rm a}$Indicative signal-to-noise ratio at 5000\,\AA\ (or 6000\,\AA\ in the case of J0140$-$0839, Q1108$-$077, J1337$+$3152, J1340$+$1106 and J2155$+$1358).\\
    \label{tab:targets}
\end{table}

The composition of the DLAs is derived
by searching for the absorption lines of elements in
their dominant ionization state (usually the neutral or
singly ionized species for DLAs) against the emission
spectrum of a more distant, background QSO. In all cases,
the QSOs were observed with either the European Southern
Observatory's (ESO) Ultraviolet and Visual Echelle
Spectrograph (UVES, \citealt{Dek00}), or the W. M. Keck Observatory's
High Resolution Echelle Spectrometer (HIRES, \citealt{Vog94}).
These spectra typically have  resolving power $R \sim 40\,000$,
and signal-to-noise ratio S/N\,$\sim 30$ per wavelength bin
($\sim 2.5$ km s$^{-1}$).
Details of the data acquisition and reduction can be
found in \citet{Coo11b}. 

For the present study, we extracted a $\pm 300$ km s$^{-1}$
window around each of the absorption lines of interest and
fitted a low order polynomial to regions deemed to be free
of absorption. The details of this continuum fit
(i.e. the polynomial order and pixels used in the continuum fit)
are stored, and are used later to derive the systematic
errors introduced by our choice of continuum.
The absorption lines targeted are listed in Table~\ref{tab:atomic};
vacuum wavelengths and oscillator strengths are from the 
compilation by \citet{Mor03} with updates by \citet{JenTri06}.

\begin{table*}
\caption{\textsc{Adopted metal line laboratory wavelengths and oscillator strengths$^{\rm a}$}}
\centering
    \begin{tabular}{lr@{.}lr@{.}lp{0.7cm}lr@{.}lr@{.}lp{0.7cm}lr@{.}lr@{.}l}
    \hline
   \multicolumn{1}{l}{Ion}
& \multicolumn{2}{l}{Wavelength}
& \multicolumn{2}{c}{$f$}
&& \multicolumn{1}{l}{Ion}
& \multicolumn{2}{l}{Wavelength}
& \multicolumn{2}{c}{$f$}
&& \multicolumn{1}{l}{Ion}
& \multicolumn{2}{l}{Wavelength}
& \multicolumn{2}{c}{$f$}\\
   \multicolumn{1}{c}{}
& \multicolumn{2}{c}{(\AA)}
& \multicolumn{2}{c}{}
&& \multicolumn{1}{c}{}
& \multicolumn{2}{c}{(\AA)}
& \multicolumn{2}{c}{}
&& \multicolumn{1}{c}{}
& \multicolumn{2}{c}{(\AA)}
& \multicolumn{2}{c}{}\\
    \hline
\SiII   & 1808&01288     & 0&00208    &&       \CoII  & 1424&7866     & 0&0109 &&       \NiII   & 1370&132          & 0&05 \\
\SII    & 1250&578         & 0&00543     &&       \CoII   & 1466&2110     & 0&031 &&      \NiII   & 1393&324     & 0&0101  \\
\SII    & 1253&805         & 0&0109       &&      \CoII   & 1480&9546     & 0&0119  &&     \NiII   & 1454&842     & 0&0323 \\
\SII    & 1259&5180       & 0&0166       &&      \CoII   & 1552&7624      & 0&0116  &&    \NiII   & 1502&1480   &  0&0133   \\
\TiII   & 1910&9538      & 0&098         &&      \CoII   & 1572&6480      & 0&0120   &&     \NiII   & 1709&6042   & 0&0324\\
\TiII   & 1910&6123      & 0&104         &&      \CoII   & 1574&5508       & 0&025  &&       \NiII   & 1741&5531   & 0&0427  \\
\CrII   & 2056&25693   & 0&1030       &&       \CoII   & 1941&2852       & 0&034  &&      \NiII   & 1751&9157   & 0&027   \\
\CrII   & 2062&23610   & 0&0759       &&      \CoII   & 2012&1664       & 0&0368   &&     \ZnII  & 2026&13709 & 0&501    \\
\CrII   & 2066&16403   & 0&0512       &&      \NiII    & 1317&217          & 0&057    &&    \ZnII  & 2062&66045 & 0&246\\
\hline
\end{tabular}
    \smallskip

\begin{flushleft}
$^{\rm a}${From Morton (2003) with updates by Jenkins \& Tripp (2006).}
\end{flushleft}
\label{tab:atomic}
\end{table*}

%%%%%%%%%%%%%%%%%%%%%%%%%%%%%%%%%%%%%%%%%%%%%%%%
\section{Analysis and Results}
\label{sec:analysis}
%%%%%%%%%%%%%%%%%%%%%%%%%%%%%%%%%%%%%%%%%%%%%%%%

The primary goal of this study is to measure the
relative abundances of a few Fe-peak
elements with atomic number in the range 22 (Ti) to 30 (Zn),
in the lowest metallicity systems known at redshifts $z \gtrsim 2$.
Of these elements, we adopt Fe as the `reference' element since
it is the most abundant of the Fe-peak elements, and its absorption
lines are detected in all of the DLAs in our sample with at least $5\sigma$
confidence.

The remaining Fe-peak elements are all less abundant than Fe,
and their absorption lines are rarely seen in the most metal-poor DLAs; 
the only reported detections in individual VMP DLAs are
those discussed here. 
The reason for such dearth of data can be appreciated 
by considering the following. 
Even the strongest \NiII\ absorption line
in Table~\ref{tab:atomic}, Ni\,{\sc ii}\,$\lambda 1317.217$,
would require a signal-to-noise ratio S/N\,\,$>\!\!110$ per
resolution element to be detected at the $5 \sigma$ 
level in a DLA with [Fe/H]\,$< -2$ and 
$\log N$(H\,{\sc i})/cm$^{-2}\,= 20.3$. 
Such a high S/N ratio is beyond current observational 
capabilities given that most VMP DLAs are found in the
spectra of faint QSOs ($V \gtrsim 18$). 
Consequently, all
measures of [Ni/Fe] in individual VMP DLAs 
reported to date refer to higher (and rarer) column density
systems than the DLA threshold
$\log N$(H\,{\sc i})/cm$^{-2}\,= 20.3$.

To circumvent this issue, we developed a new type of stacking technique
to measure, or place limits on,
the ratios of several Fe-peak elements in the DLAs in our sample.

\subsection{Stacking Technique}
\label{sec:stacktech}

The stacking technique that we now describe can be used to combine
either: (i) multiple transitions of a given ion for a single DLA; or
(ii) all transitions of a given ion in \emph{all} DLAs (assuming
that Fe-peak element ratios do not vary significantly between
different VMP DLAs).
In individual DLAs where we
detect at least one absorption line from an Fe-peak element
(other than Fe),
we used the former approach and stacked all transitions of that ion
covered by our spectrum of the DLA, whether detected or not. 
This has the effect of utilising all of the information available
to improve the precision of the ion column density determination
over that achievable from the analysis of only the absorption lines
that have been detected. For DLAs where a given ion has not been detected, 
we adopted the latter approach and produced a `population' stack.

One advantage of working with weak, undetected absorption lines
is that they fall on the linear part of the curve of growth\footnote{This 
statement does not necessarily apply to low resolution spectra, but is
valid at the high spectral resolution of our echelle observations.}.
In this regime, one can simply relate the rest-frame equivalent
width of an absorption line to the column density (cm$^{-2}$)
of element X in ionization state \textsc{n} by:

\begin{equation}
\label{eq:ewn}
{\rm EW}({\rm X\,\textsc{n}}) = \frac{N({\rm X\,\textsc{n}})\,f\,\lambda^2}{1.13\times10^{20}},
\end{equation}

\noindent where $f$ and $\lambda$ are the oscillator strength and the rest
wavelength of the atomic transition (in \AA) respectively;
values relevant to this work
are collected in Table~\ref{tab:atomic}.
This technique has the clear
advantage of being independent of the cloud modelling, 
and the spectral resolution. 
Given that in all cases we have at least one 
unsaturated \FeII\ absorption line from which
the column density of singly ionized iron,
$N$(\FeII), can be deduced with confidence,
equation~(\ref{eq:ewn}) can be rearranged as follows:

\begin{equation}
\label{eq:nxnfe}
\bigg(\frac{N({\rm X\,\textsc{n}})}{N({\rm Fe\,\textsc{ii}})}\bigg)_{\rm pp} = {\rm EW}_{\rm pp}({\rm X\,\textsc{n}})\frac{1.13\times10^{20}}{N({\rm Fe\,\textsc{ii}})\,f\,\lambda^2},
\end{equation}

\noindent where EW$_{\rm pp}$ is the equivalent width per pixel.
Thus, for each pixel in the spectrum where we see absorption
by X\,\textsc{n}, the rest-frame equivalent
width of this pixel alone [weighted by the fraction on the
right side of equation~(\ref{eq:nxnfe})] represents the
fractional contribution to the total column density
ratio of species X\,\textsc{n} relative to \FeII.
If the width of each pixel is measured in km s$^{-1}$,
the quantity on the left side of equation~(\ref{eq:nxnfe})
has the units km$^{-1}$ s. Integrating
equation~(\ref{eq:nxnfe}) over all pixels
where the absorption takes place then gives the 
final value of $N({\rm X\,\textsc{n}})/N({\rm Fe\,\textsc{ii}})$.

For a given DLA, equation~(\ref{eq:nxnfe}) can be used to
create a `stacked' spectrum for \emph{all} lines of a
given species. As an example, the suite of \NiII\ lines
available in many of the DLA spectra can be combined to
provide a single, more precise measure of the \NiII/\FeII\
ratio. To do this, we first convert the spectrum in
the vicinity of each transition to a common velocity scale.
The zero point of the velocity scale is established 
from the profiles of the \FeII\ lines available.
We then calculate the rest-frame equivalent width for each
pixel for all transitions and use equation~(\ref{eq:nxnfe})
to convert this to a column density ratio. Finally, we
determine the weighted mean of all transitions for a given
species, where the weights correspond to the inverse variance
of each pixel. In this way, we ensure that the final stack
has the highest achievable S/N. 
We stress that
this technique can be implemented for a single DLA
without any assumptions. Therefore, for DLAs where we
detect at least one transition of a given Fe-peak element
(apart from Fe itself),
we will use this technique
to stack all the available transitions in that DLA spectrum
and obtain the value of 
$N({\rm X\,\textsc{n}})/N({\rm Fe\,\textsc{ii}})$
appropriate to that DLA.

The stacking method just described
can only be applied to the few cases 
where we do detect at least one absorption
line of a given Fe-peak element. However,
for most VMP DLAs in our survey, even stacking
all the (undetected) absorption lines of a given ion
in the spectrum will not improve the S/N sufficiently to
lead to a meaningful measure of 
$N({\rm X\,\textsc{n}})/N({\rm Fe\,\textsc{ii}})$.
In these circumstances, however, we can
still apply the above method to construct a 
stacked spectrum of  all lines of a given
ion for \emph{all} DLAs with non-detections.
In this way, we can derive an 
averaged measure of $N({\rm X\,\textsc{n}})/N({\rm Fe\,\textsc{ii}})$
(or an upper limit to this ratio)
for the population of VMP DLAs in our sample.
Of course, such an average value does
not tell us about the \textit{spread} of values of
$N$(X\,\textsc{n})/$N$(Fe\,\textsc{ii})
within the population of DLAs, but it is
nevertheless instructive, given the
relatively small scatter in the ratios
of most Fe-peak elements exhibited by Galactic
halo stars of comparable metallicity \citep{Cay04,Lai08,Yon12}.

There are two potential systematics that may introduce a bias
when stacking the absorption lines for a selection of DLAs with
a variety of metallicities and \HI\ column densities.
The first concern is that certain species may be depleted
(relative to Fe) onto dust grains, 
an effect which is known to depend
on metallicity. Fortunately, when the DLA metallicity is
[Fe/H]\,$\lesssim-2.0$, dust depletions become
unimportant \citep{Pet97,Ake05}. Furthermore, 
the elements that we consider here have roughly the same
affinity for dust (excluding Zn; see our discussion in
Secion~\ref{sec:starsZn}) so their \emph{relative} dust
depletions are negligible. The other
concern is that the radiation field to which the gas
is exposed may vary from DLA to DLA, depending
among other factors on the \HI\ column density.
However, for the ions of the elements
under consideration here, which are the
dominant ions in neutral gas, 
the \emph{differential} ionization corrections required to convert
the measured $N({\rm X\,\textsc{n}})/N({\rm Fe\,\textsc{ii}})$
to the element abundance ratio X/Fe are minimal and
can be safely neglected (see e.g. \citealt{Vla01,Coo10}).

Thus, in what follows we assume
$N$(X\,\textsc{n})/$N$(\FeII) = $N$(X)/$N$(Fe).
Comparing the above ratio to the solar value then gives the 
quantity of interest, 
[X/Fe]\,$\equiv\log (N{\rm (X)}/N{\rm (Fe)}) - \log {\rm (X/Fe)}_{\odot}$.
We adopt the \citet{Asp09} solar abundance scale,
taking either the photospheric, meteoritic, or an average
of the two determinations following
\citet{LodPlaGai09}. The solar abundances
adopted in this work are listed in Table~\ref{tab:solabu}.

\begin{table}
\caption{\textsc{Adopted solar abundances$^{\rm a}$}}
\centering
    \begin{tabular}{lcp{0.5cm}lc}
    \hline
   \multicolumn{1}{l}{X}
& \multicolumn{1}{c}{log(X/H)$_{\odot}$}
&& \multicolumn{1}{l}{X}
& \multicolumn{1}{c}{log(X/H)$_{\odot}$}\\
    \hline
Si  & $-4.49$ & &   Fe  & $-4.53$ \\
S   & $-4.86$ & &   Co & $-7.07$ \\
Ti  & $-7.09$ & &   Ni   & $-5.79$ \\
Cr  & $-6.36$ & &  Zn  & $-7.37$  \\
\hline
\end{tabular}
    \smallskip

\begin{flushleft}
$^{\rm a}${From Asplund et al. (2009).}
\end{flushleft}
\label{tab:solabu}
\end{table}

%%%%%%%%%%%%
% FIGURE 1 %
%%%%%%%%%%%%
\begin{figure}
  \centering
 {\includegraphics[angle=0,width=80mm]{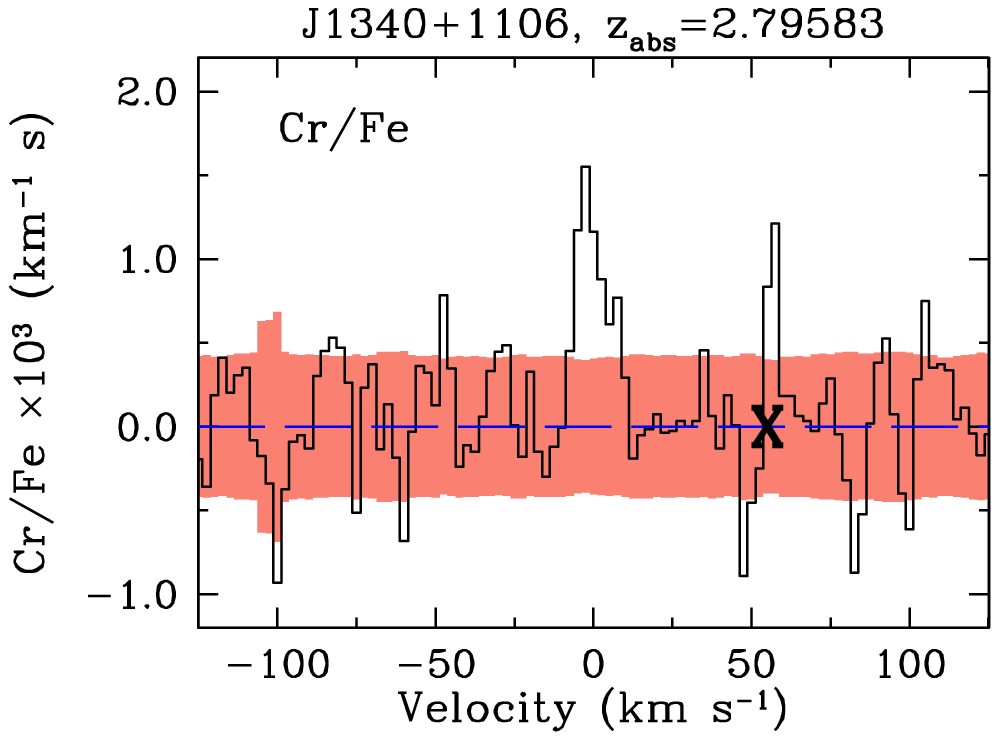}}\\
 {\vspace{0.4cm}}
 {\includegraphics[angle=0,width=80mm]{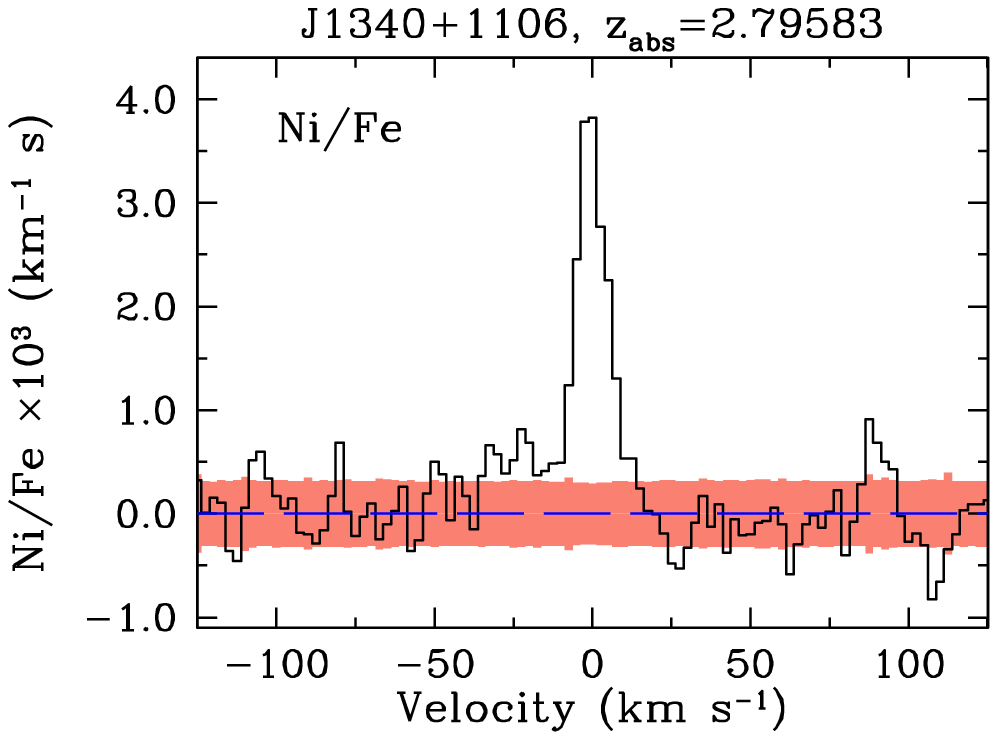}}\\
  \caption{
The stacked spectrum (black histogram) that is used to measure
the Cr/Fe and Ni/Fe ratios (top and bottom panel respectively)
in the DLA at $z_{\rm abs}=2.79583$ towards J1340$+$1106.
The shaded red bands show the $\pm 1\sigma$ uncertainty, whereas the
long-dashed blue lines indicate the levels for no \CrII\ and \NiII\ absorption. 
The black `X' at $+55$ km s$^{-1}$ in the top panel marks
an unrelated feature.
  }
  \label{fig:testcase}
\end{figure}

The most likely source of systematic error is in the
choice of the continuum around each absorption line,
which is especially important when dealing with weak
lines. To estimate the additional uncertainty (and bias)
that is introduced by our placement of the continuum level,
we adopted a Monte Carlo approach similar
to that described by \citet{PetCoo12}. 
We used the measured line equivalent widths (or upper limits) 
to generate a series of line profiles for each of the transitions
used in the stack. We then combined these profiles with the
stacking technique described above---first without any
continuum adjustments and then again with the identical
continuum adjustments applied to the real data (see Sec.~\ref{sec:sam}).
This exercise was repeated 1000 times, and the deviations
between the equivalent widths derived with and without the
continuum adjustments were taken to be a measure
of the systematic uncertainty.
In general, we found that the systematic errors are of
the same order as the random errors.
Perhaps more importantly, we also verified that
in all of the cases described below there were \emph{no
significant systematic biases} introduced by our 
estimate of the continuum level.

\subsection{J1340$+$1106 -- A test case}

We tested the validity of this technique
by applying it to one of the few clear detections
of \CrII\ and \NiII\ absorption in our sample.
With a relatively high column density
$N$(\HI)\,$= 1 \times 10^{21}$\,cm$^{-2}$,
the DLA at redshift $z_{\rm abs}=2.79583$ toward J1340$+$1106
shows an unusually high number of metal lines
for its low metallicity [Fe/H]\,$= -2.15$ \citep{Coo11b}.
From the analysis of 36 absorption lines from 11
ions dominant in H\,{\sc i} regions,
\citet{Coo11b} derived a `cloud model' which 
successfully reproduced the line profiles.
This model was used to derive the ion column densities,
including those of \CrII\ and \NiII.

In Fig.~\ref{fig:testcase}, we have reproduced
the stacks of \CrII\ and \NiII\ lines
derived from equation~(\ref{eq:nxnfe}),
showing clear detections near relative velocity 
 $v=0$ km s$^{-1}$. The total \CrII/\FeII\ and \NiII/\FeII\
ratios are obtained by integrating these profiles over
the velocity range of the absorption, defined by
the extent of the stronger absorption lines of \FeII.
We find [Cr/Fe]\,$= 0.08 \pm 0.11$
and [Ni/Fe]\,$= 0.03 \pm 0.03$. These values 
compare well with [Cr/Fe]\,$= 0.13 \pm 0.11$
and [Ni/Fe]\,$= 0.02 \pm 0.03$
deduced by \citet{Coo11b} from profile fitting
of the absorption lines.

\subsection{Individual systems with detections}
\label{sec:detect}

We now apply the stacking technique described in
Section~\ref{sec:stacktech} to the five DLAs
in our sample that: (i) either show absorption by at least one 
Fe-peak element (other than Fe), or (ii) where the data
are of sufficient quality to allow a useful upper limit
to be set. For this to be the case, these VMP DLAs
either have a large \HI\ column density, or their
spectra have been recorded at an unusually high S/N, or both.
Even in these most favourable cases, however,
the only Fe-peak elements detected are Ni and Cr, 
the two most abundant after Fe.
In Table~\ref{tab:summary} we list five detections
(four for \NiII\ and one for \CrII) and three upper limits (one for \NiII\
and two for \CrII). 
For each system, we give [Fe/H], 
the significance of detections of \NiII\ and \CrII, 
and the number of transitions used in the stacked 
spectrum. In all cases,
we integrated the stacked profile over the velocity
range exhibited by the stronger \FeII\ absorption lines,
which is provided in column 6 of Table~\ref{tab:summary}.

The stacked
profiles for these five DLAs are shown in 
Fig.~\ref{fig:testcase} and Fig.~\ref{fig:Dstack}.
All of the right panels in Fig.~\ref{fig:Dstack}
have significant features (see penultimate column of
Table~\ref{tab:summary}). 
In the lower-left and middle-left panels of the figure,
we see no discernible absorption and deduce corresponding $3\sigma$
upper limits to the ratios Ni/Fe and Cr/Fe. 
The top-left panel shows a 
feature that is `detected' at the $3\sigma$ level from which 
we conservatively derive a  $3 \sigma$ upper limit
to [Cr/Fe].

\subsection{Systems with non-detections}
\label{sec:nodetect}

The remaining 16 DLAs in our sample do not
show individual absorption lines from any Fe-peak
element besides Fe; these systems are now
used to produce a mean `population' stack. 
As discussed earlier, in generating this stack
we assume that VMP DLAs do not exhibit a wide range of
[X/Fe] values over the metallicity range probed
here, $-3.5 \leq {\rm [Fe/H]} \leq -2.0$.
Cases with individual detections described
in Section~\ref{sec:detect} were \emph{not}
included in the stacks.
By inspecting the \FeII\ absorption
lines for all systems in the stack, we determined
that most absorption occurs over the velocity interval
$\Delta v = \pm 25$ km s$^{-1}$ from the 
reference redshift. We then confirmed that all
of the undetected absorption lines used in
the stack contained no significant absorption within
this velocity interval (i.e. all pixels are within
$\pm 2\sigma$ of the continuum level in this velocity interval).
Furthermore, to be certain that we are not including
spurious contributions from coincident, unrelated
absorption, we rejected the high and low values for
each pixel in the stack. This procedure ensures that
we suffer minimal contamination. 

\begin{figure*}
  \centering
 {\hspace{-0.25cm} \includegraphics[angle=0,width=70mm]{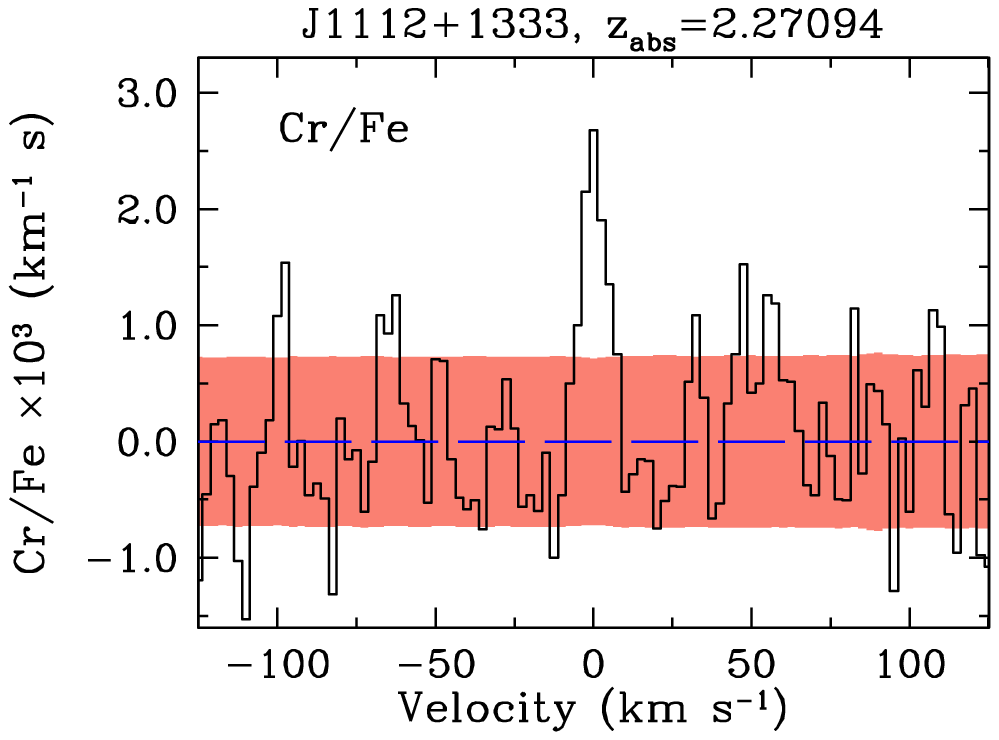}}
 {\hspace{0.35cm} \includegraphics[angle=0,width=70mm]{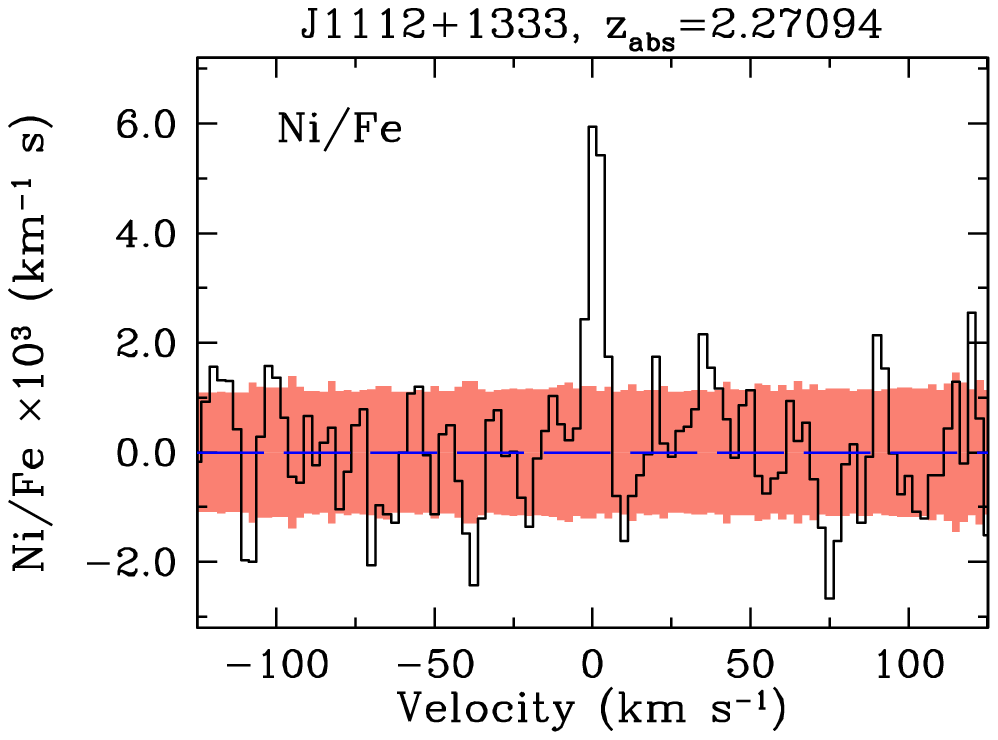}}\\
 {\vspace{0.4cm}}
 {\hspace{-0.25cm} \includegraphics[angle=0,width=70mm]{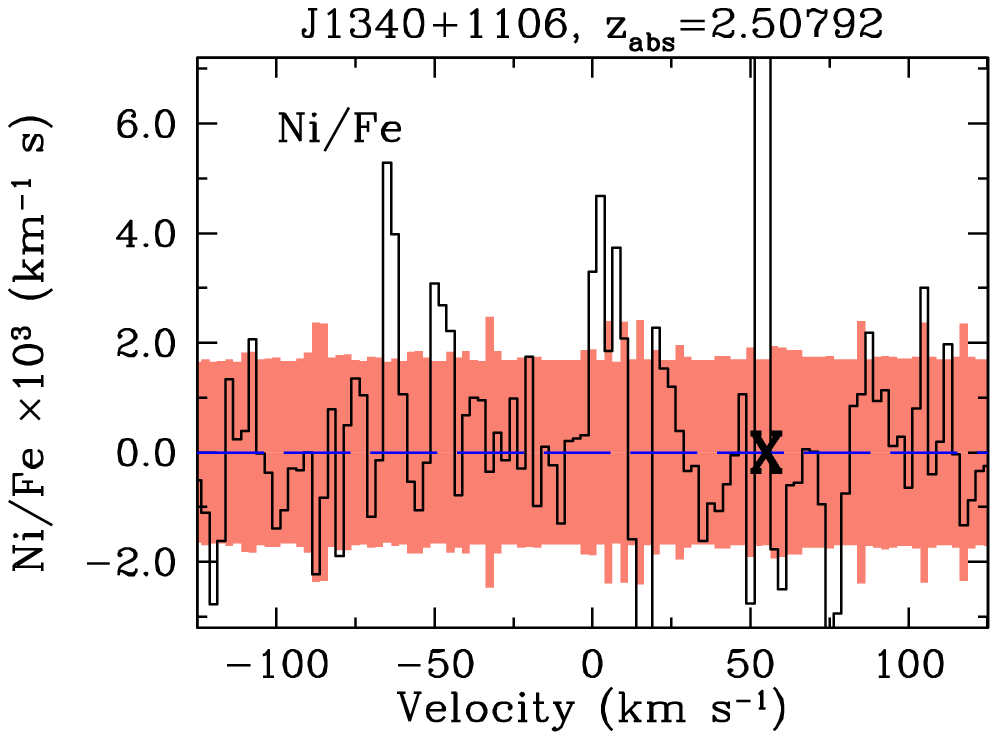}}
 {\hspace{0.35cm} \includegraphics[angle=0,width=70mm]{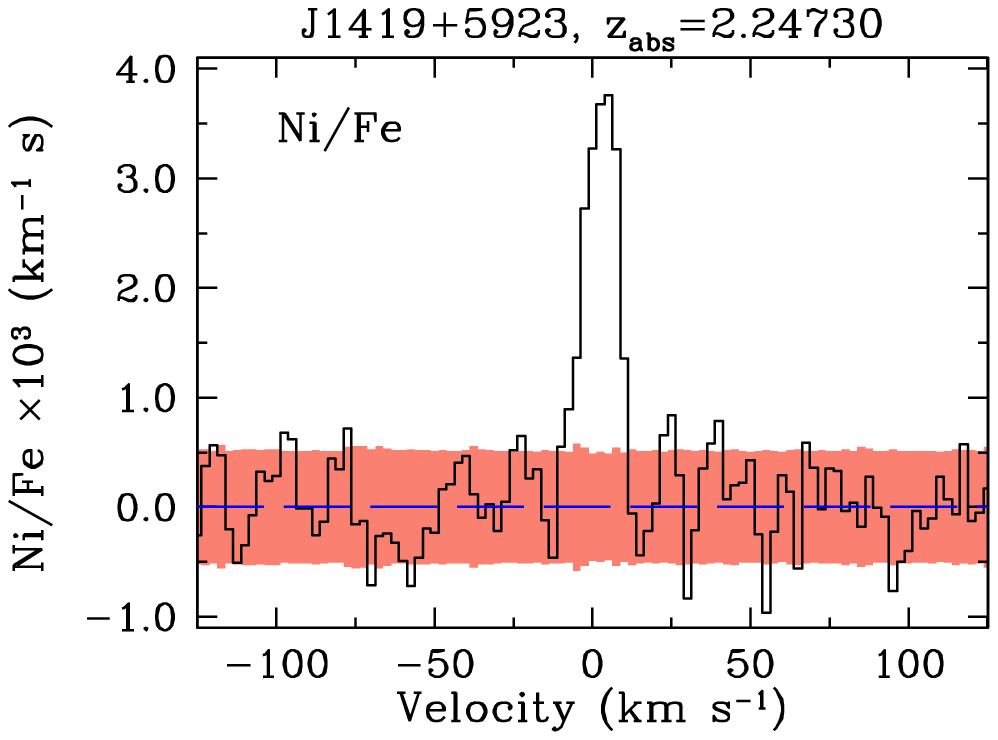}}\\
 {\vspace{0.4cm}}
 {\hspace{-0.25cm} \includegraphics[angle=0,width=70mm]{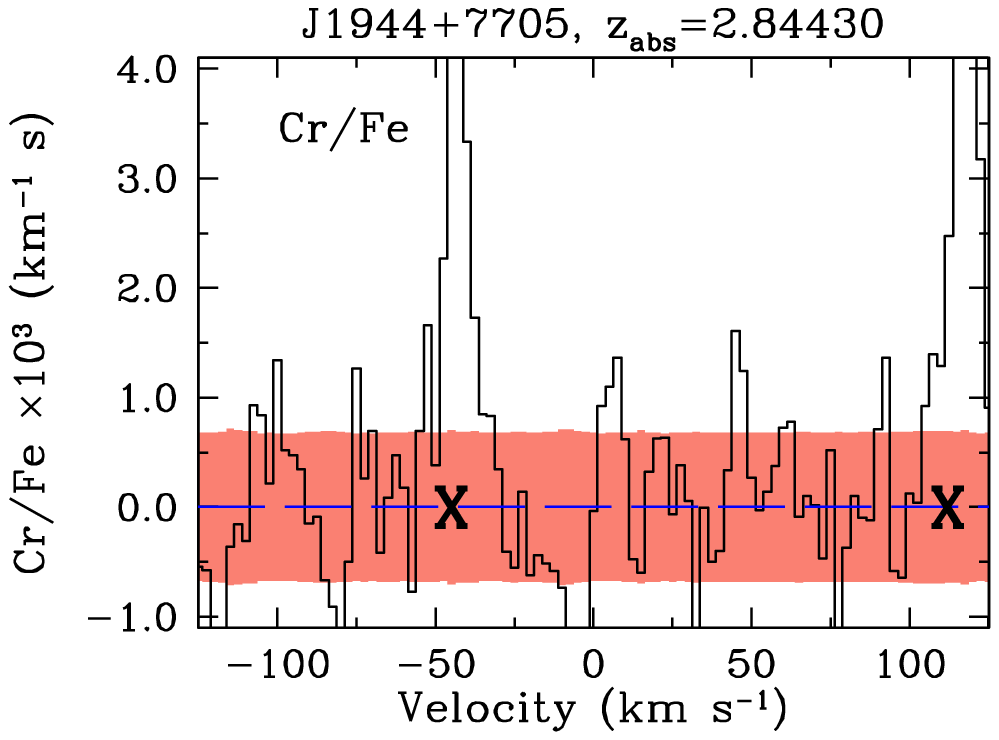}}
 {\hspace{0.35cm} \includegraphics[angle=0,width=70mm]{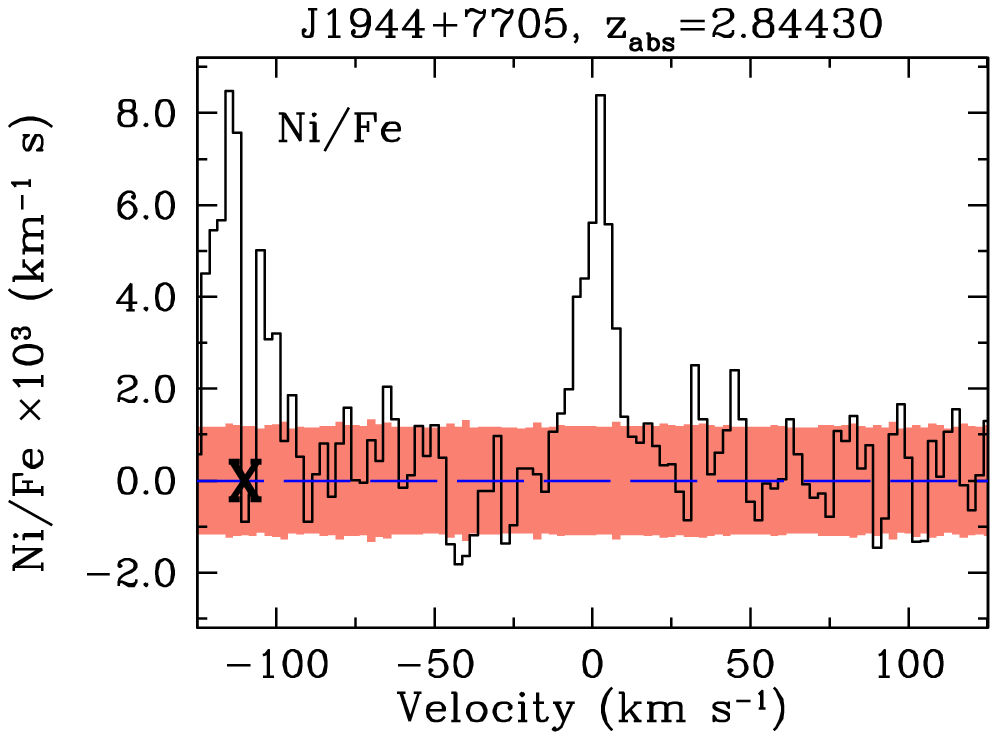}}\\
  \caption{
Stacked spectra (black histograms) for individual DLAs in which at least one
absorption line from an Fe-peak element (other than Fe) was detected,
or a useful upper limit could be determined. 
The red shaded regions correspond to the $\pm 1\sigma$
random errors. The three panels on the right show significant detections
of \NiII,  while from the three panels on the left $3\sigma$ upper limits
to [Cr/Fe] and [Ni/Fe] can be determined.
Black `X' symbols denote unrelated absorption features.
  }
  \label{fig:Dstack}
\end{figure*}

\begin{figure*}
  \centering
 {\hspace{-0.25cm} \includegraphics[angle=0,width=70mm]{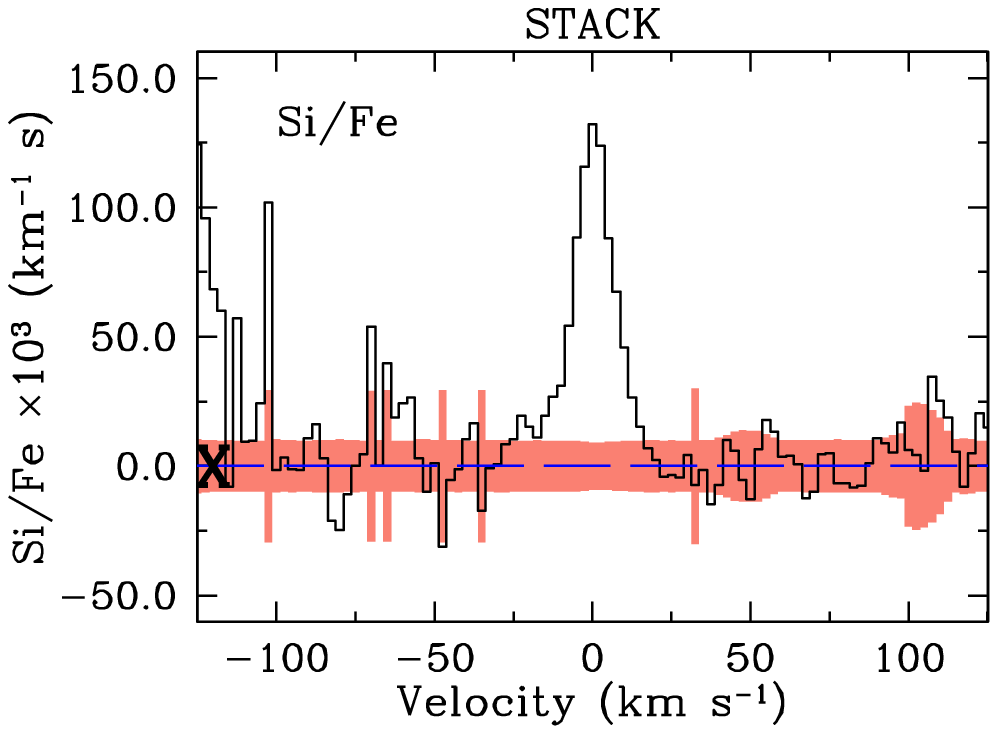}}
 {\hspace{0.35cm} \includegraphics[angle=0,width=70mm]{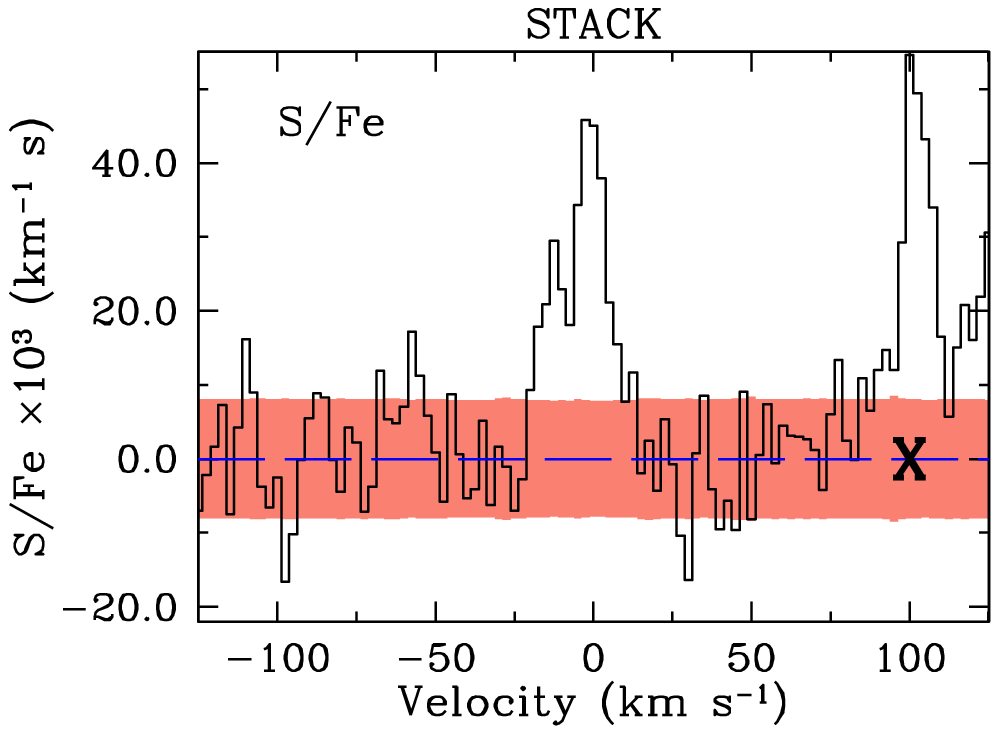}}\\
 {\vspace{0.1cm}}
 {\hspace{-0.25cm} \includegraphics[angle=0,width=70mm]{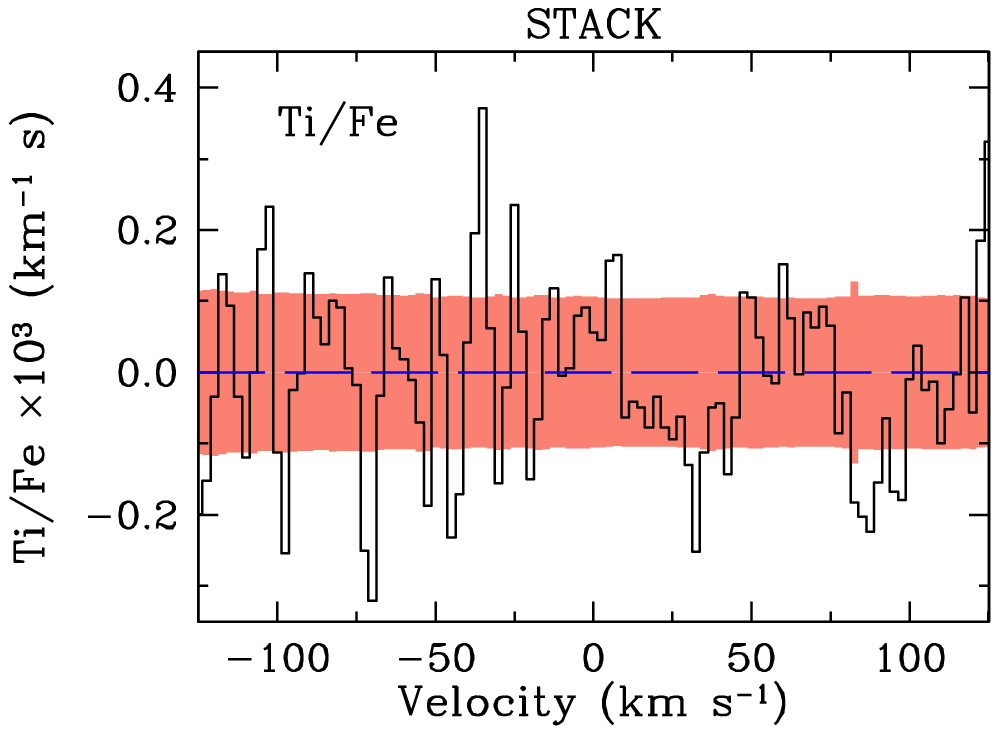}}
 {\hspace{0.35cm} \includegraphics[angle=0,width=70mm]{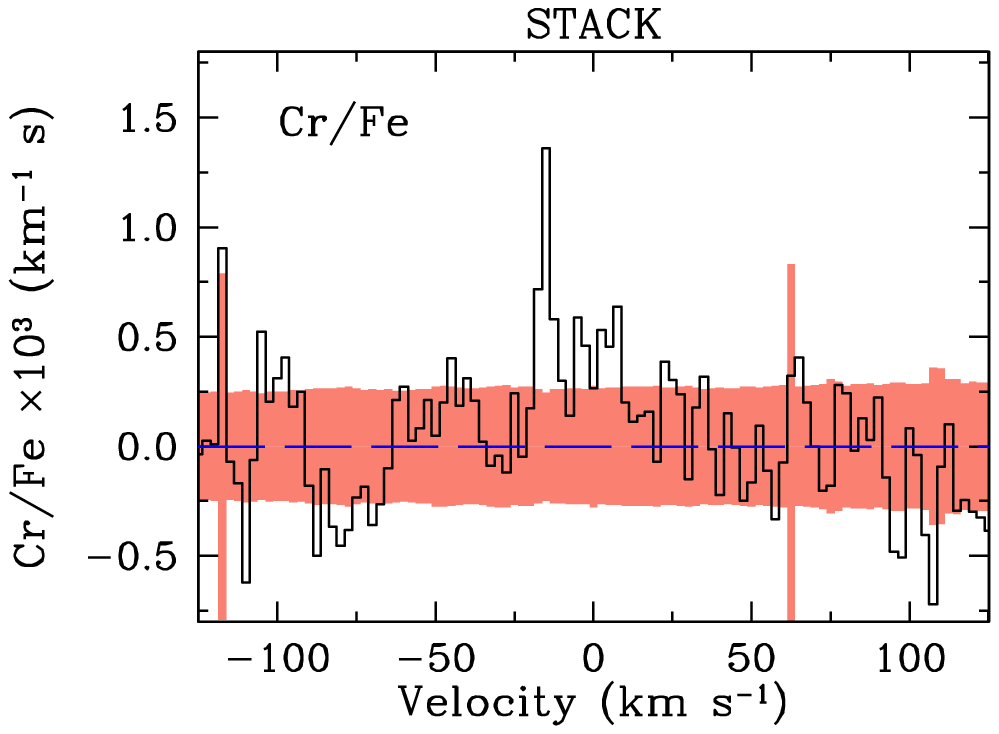}}\\
 {\vspace{0.1cm}}
 {\hspace{-0.25cm} \includegraphics[angle=0,width=70mm]{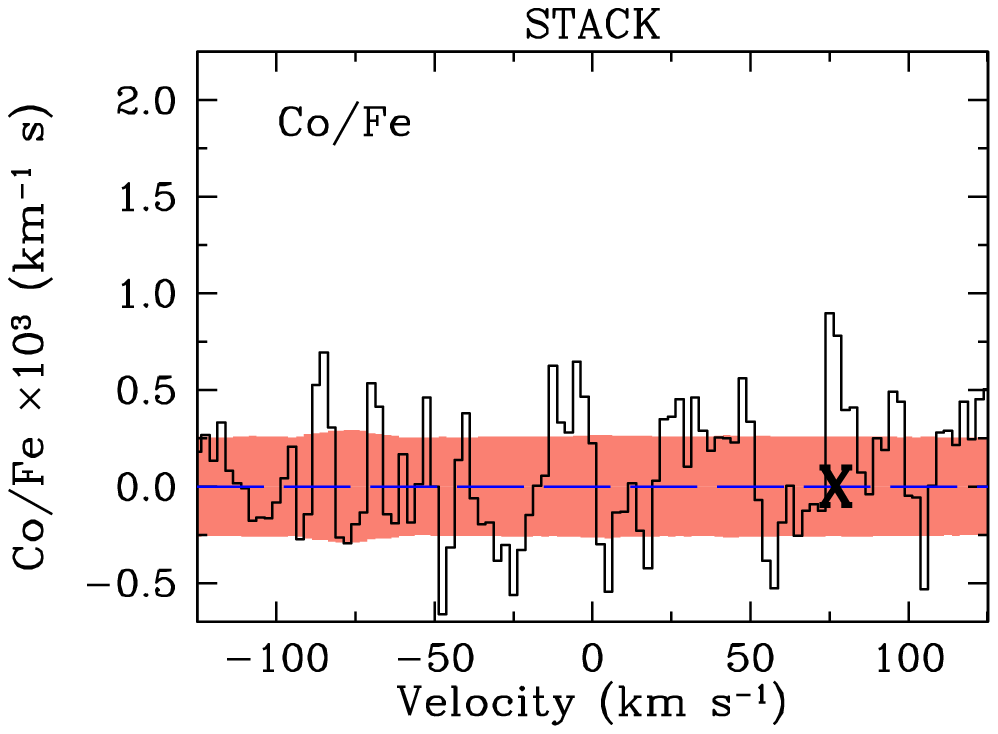}}
 {\hspace{0.35cm} \includegraphics[angle=0,width=70mm]{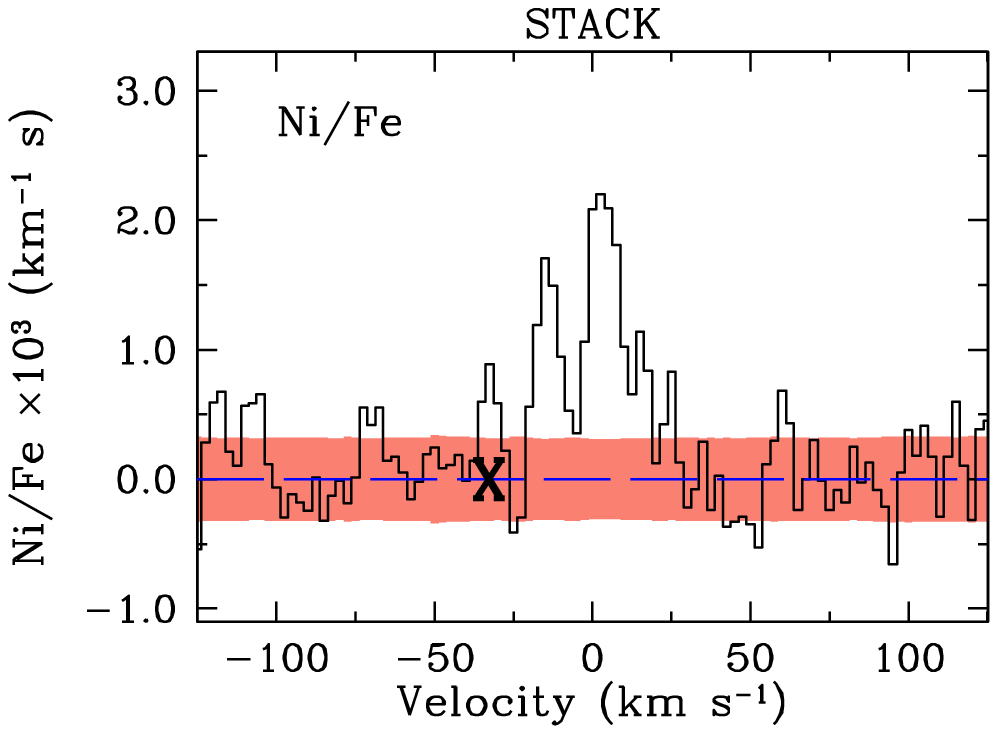}}\\
 {\vspace{0.1cm}}
 {\includegraphics[angle=0,width=70mm]{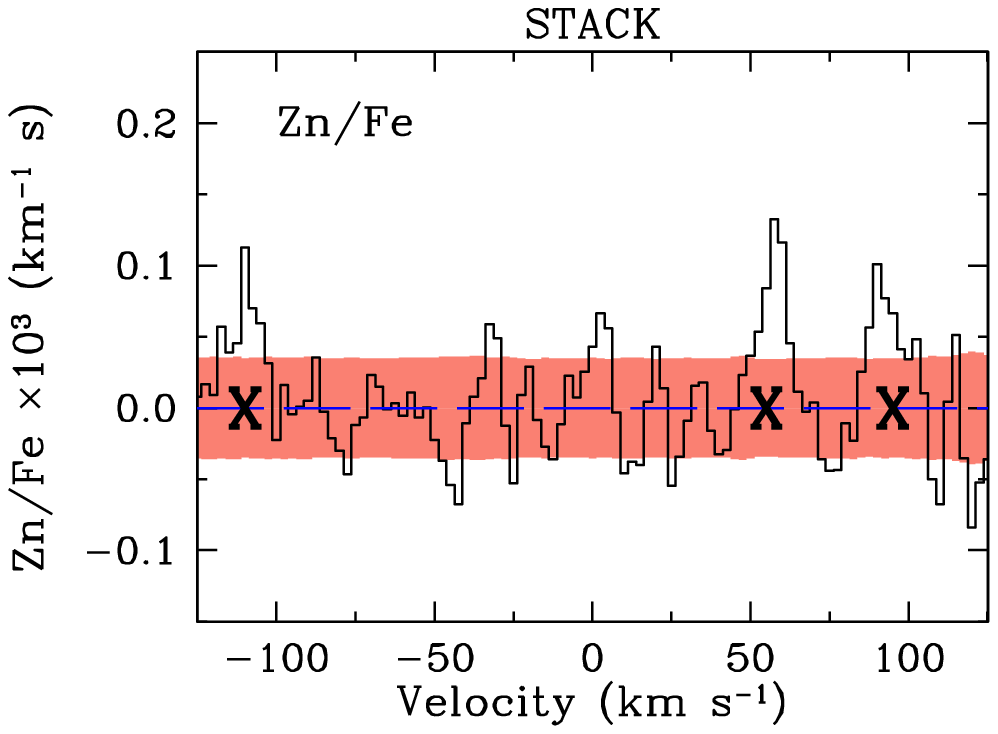}}
  \caption{
Stacked spectra (black histograms) for DLAs without individual detections
of Fe-peak elements. The red shaded regions correspond to the $\pm 1\sigma$
random errors. We have clear detections for Si/Fe, S/Fe, Cr/Fe, and Ni/Fe
and $3\sigma$ upper limits for Ti/Fe, Co/Fe, and Zn/Fe. Black `X' 
symbols denote unrelated absorption features.
  }
  \label{fig:Nstack}
\end{figure*}

The results of these stacks are reproduced in Fig.~\ref{fig:Nstack}.
Due to the wider variety of line
profile shapes that have been used in the stack
(which in some cases contain multiple components
and extend to $\pm 25$ km s$^{-1}$), the features
appear more broadened than those shown in
Fig.~\ref{fig:Dstack} which typically exhibit a single
dominant absorption component.
The corresponding values of [X/Fe] are 
listed in the fourth column of
Table~\ref{tab:summary}. 
The metallicity [Fe/H] to which these
values refer is not the same for 
every element X considered, because 
not all of the 16 DLAs contribute the same number
of absorption lines to each stack. 
The values of [Fe/H] and their ranges
listed in the fifth column of  Table ~\ref{tab:summary}
under the heading `Stacked systems with non-detections'
were determined using a weighted bootstrap method
on the array of Fe/H values for each stack,
with $N$(\FeII)\,$\times {\rm (S/N)}^2$ as the 
weight on each line used in the stack.

In order to check the consistency of our results,
we also produced two stacks of lines 
from $\alpha$-capture elements, one
for the weak \SiII\,$\lambda 1808$ line, and the
other for the \SII\ triplet
at $\lambda\lambda1250,\,1253,\,1259$.
As can be seen from Table~\ref{tab:summary},
we find [Si/Fe]\,$= +0.31 \pm 0.04 \pm 0.05$
(random and systematic error respectively)
and [S/Fe]\,$ = +0.25 \pm 0.05 \pm 0.06$.
The modest enhancement by a factor of $\sim 2$
of these two $\alpha$-capture elements relative to Fe
is consistent with the mean value
$[\langle {\rm Si/Fe} \rangle ] = +0.32 \pm 0.09$
deduced by \citet{Coo11b} in the same
metallicity interval.

%%%%%%%%%%%%
% FIGURE 2 %
%%%%%%%%%%%%

%%%%%%%%%%%%
% FIGURE 3 %
%%%%%%%%%%%%

\subsection{Other systems in the literature}

To our knowledge, this is the first detailed investigation into the
Fe-peak element ratios in very metal-poor DLAs. This is perhaps not
surprising, given that there are just a few VMP DLAs that we could
find in the literature with column density measurements of
Fe-peak elements:
(1) the DLA with \NHI/cm$^{-2}=21.41\pm0.08$ at $z_{\rm abs}=3.3901$
    towards Q0000$-$2619 \citep{Mol00};
(2) the $z_{\rm abs}=3.915$ DLA towards Q0255$+$00 with
    \NHI/cm$^{-2}=21.30\pm0.05$ \citep{Pro01};
(3) the system at $z_{\rm abs}=2.615$ with \NHI/cm$^{-2}=21.30\pm0.10$
    towards Q2348$-$01 \citep{Pro01}; and
(4) the DLA towards Q2348$-$1444 at $z_{\rm abs}=2.279$ with
    \NHI/cm$^{-2}=20.59\pm0.08$ \citep{Des07}. 
Other relevant details for these absorbers are collected in
Table~\ref{tab:summary}, under the heading `Literature sample'.
All literature values have been adjusted to
the solar abundance scale adopted here.

%%%%%%%%%%%%%%
%  TABLE 3
%%%%%%%%%%%%%%

\begin{table*}
\caption{\textsc{Summary of stacked abundances}}
\centering
    \begin{tabular}{lccccccc}
    \hline
   \multicolumn{1}{l}{Name}
& \multicolumn{1}{c}{$z_{\rm abs}$}
& \multicolumn{1}{c}{X}
& \multicolumn{1}{c}{[X/Fe]$^{\rm a}$}
& \multicolumn{1}{c}{[Fe/H]}
& \multicolumn{1}{c}{$\Delta v$}
& \multicolumn{1}{c}{Confidence}
& \multicolumn{1}{c}{Number}\\
   \multicolumn{1}{l}{}
& \multicolumn{1}{c}{}
& \multicolumn{1}{c}{}
& \multicolumn{1}{c}{}
& \multicolumn{1}{c}{}
& \multicolumn{1}{c}{km s$^{-1}$}
& \multicolumn{1}{c}{}
& \multicolumn{1}{c}{of lines}\\
    \hline
\multicolumn{8}{l}{\textbf{Individual systems with detections}}\\
J$1112+1333$          & $2.27094$ & Cr &   $\le+0.12$                           &  $-2.33\pm0.06$           &      $\pm10$       &  $3\sigma$    &   2  \\
J$1112+1333$          & $2.27094$ & Ni &   $-0.19\pm0.12\pm0.06$   &  $-2.33\pm0.06$           &      $\pm10$       &  $4\sigma$    &   6  \\
J$1340+1106$          & $2.50792$ & Ni &   $\le+0.02$                            &  $-2.07\pm0.05$          &      $^{+15}_{-30}$        &  $3\sigma$    &   5  \\
J$1340+1106$          & $2.79583$ & Cr &   $+0.08\pm0.11\pm0.10$  &  $-2.15\pm0.06$           &      $^{+15}_{-30}$       &  $4\sigma$    &   2  \\
J$1340+1106$          & $2.79583$ & Ni &   $+0.03\pm0.03\pm0.02$  &  $-2.15\pm0.06$           &      $^{+15}_{-30}$       &  $17\sigma$   &   7  \\
J$1419+5923$          & $2.24730$ & Ni &   $-0.04\pm0.05\pm0.03$   &  $-2.07\pm0.10$           &      $\pm20$       &  $10\sigma$    &   7  \\
J$1944+7705$          & $2.84430$ & Cr &   $\le+0.16$                           &  $-2.50\pm0.06$           &      $\pm15$       &  $3\sigma$    &   2  \\
J$1944+7705$          & $2.84430$ & Ni &   $+0.25\pm0.05\pm0.03$   &  $-2.50\pm0.06$          &      $\pm15$        &  $9\sigma$    &   7  \\
\hline
\multicolumn{8}{l}{\textbf{Stacked systems with non-detections}}\\
STACK                 &    $-$    & Si &   $+0.31\pm0.04\pm0.05$  &  $-2.33^{+0.11}_{-0.15}$  &   $\pm25$          &  $13\sigma$   &   6  \\
STACK                 &    $-$    & S  &   $+0.25\pm0.05\pm0.06$  &  $-2.55^{+0.08}_{-0.10}$  &    $\pm25$         &  $6\sigma$    &  14  \\
STACK                 &    $-$    & Ti &   $\le+0.16$             &  $-2.19^{+0.09}_{-0.11}$                 &   $\pm25$          &  $3\sigma$    &  16  \\
STACK                 &    $-$    & Cr &   $+0.11\pm0.08\pm0.09$  &  $-2.27^{+0.14}_{-0.21}$  &   $\pm25$          &  $5\sigma$    &   8  \\
STACK                 &    $-$    & Co &   $\le+0.53$  &  $-2.17^{+0.04}_{-0.05}$                          &  $\pm25$            &  $3\sigma$    &   85  \\
STACK                 &    $-$    & Ni &   $-0.03\pm0.04\pm0.03$  &  $-2.19^{+0.08}_{-0.10}$  &    $\pm25$          &  $12\sigma$   &  52  \\
STACK                 &    $-$    & Zn &   $\le-0.04$             &  $-2.32^{+0.10}_{-0.13}$               &    $\pm25$         &  $3\sigma$    &  11  \\
\hline
\multicolumn{8}{l}{\textbf{Literature sample}}\\
Q$0000-2619^{\rm b}$  & $3.39010$ & Cr &   $+0.05\pm0.04$         &  $-2.01\pm0.09$           &     $-$           &     $-$       &  1  \\
Q$0000-2619^{\rm b}$  & $3.39010$ & Ni &   $-0.22\pm0.04$         &  $-2.01\pm0.09$              &     $-$        &     $-$       &  1  \\
Q$0000-2619^{\rm b}$  & $3.39010$ & Zn &   $-0.02\pm0.06$         &  $-2.01\pm0.09$           &     $-$           &     $-$       &  1  \\
Q$0255+00^{\rm c}$    & $3.91500$ & Ni &   $-0.22\pm0.10$         &  $-2.02\pm0.10$           &     $-$           &     $-$       &  3  \\
Q$2348-01^{\rm c}$    & $2.61500$ & Cr &   $-0.07\pm0.11$         &  $-2.20\pm0.13$            &     $-$          &     $-$       &  2  \\
Q$2348-01^{\rm c}$    & $2.61500$ & Ni &   $-0.12\pm0.11$         &  $-2.20\pm0.13$            &     $-$          &     $-$       &  2  \\
Q$2348-01^{\rm c}$    & $2.61500$ & Zn &   $\le0.14$              &  $-2.20\pm0.13$              &     $-$        &     $-$       &  1  \\
Q$2348-1444^{\rm d}$  & $2.27900$ & Cr &   $+0.29\pm0.12$         &  $-2.21\pm0.09$              &     $-$        &     $-$       &  $\le3$  \\
Q$2348-1444^{\rm d}$  & $2.27900$ & Ni &   $\le-0.28$             &  $-2.21\pm0.09$                &     $-$      &     $-$       &  1  \\
Q$2348-1444^{\rm d}$  & $2.27900$ & Zn &   $\le+0.28$             &  $-2.21\pm0.09$              &     $-$        &     $-$       &  1  \\
\hline
\end{tabular}
    \smallskip

\begin{flushleft}
$^{\rm a}${For the cases where two errors are given, they refer to random 
and systematic errors respectively.}\\
$^{\rm b}${\citet{Mol00}.}\\
$^{\rm c}${\citet{Pro01}.}\\
$^{\rm d}${\citet{Des07}.}
\end{flushleft}
\label{tab:summary}
\end{table*}

%%%%%%%%%%%%%%%%%%%%%%%%%%%%%%%%%%%%%%%%%%%%%%%%%%%%%%%%%
%%%%%%%%%%%%%%%%%%%%%%%%%%%%%%%%%%%%%%%%%%%%%%%%%%%%%%%%%

\section{Comparison with low-metallicity stars}
\label{sec:stars}

Our knowledge of the Fe-peak elements
in the VMP regime ([Fe/H]\,$\le -2.0$) is currently
based primarily on observations of the most metal-poor stars
in the halo of our Galaxy and its companions. 
VMP DLAs offer a new probe
to verify independently the validity of the stellar 
trends with metallicity.
One advantage of using DLAs to measure
relative element abundances is that the
derived ratios are model independent; in metal-poor
stars, systematic uncertainties in the abundances may arise for
some absorption lines that are modelled in one-dimension,
or if the assumption of local thermodynamic equilibrium (LTE)
is not valid.

%%%%%%%%%%%%
% FIGURE 4 %
%%%%%%%%%%%%
\begin{figure}
  \centering
  \includegraphics[angle=0,width=80mm]{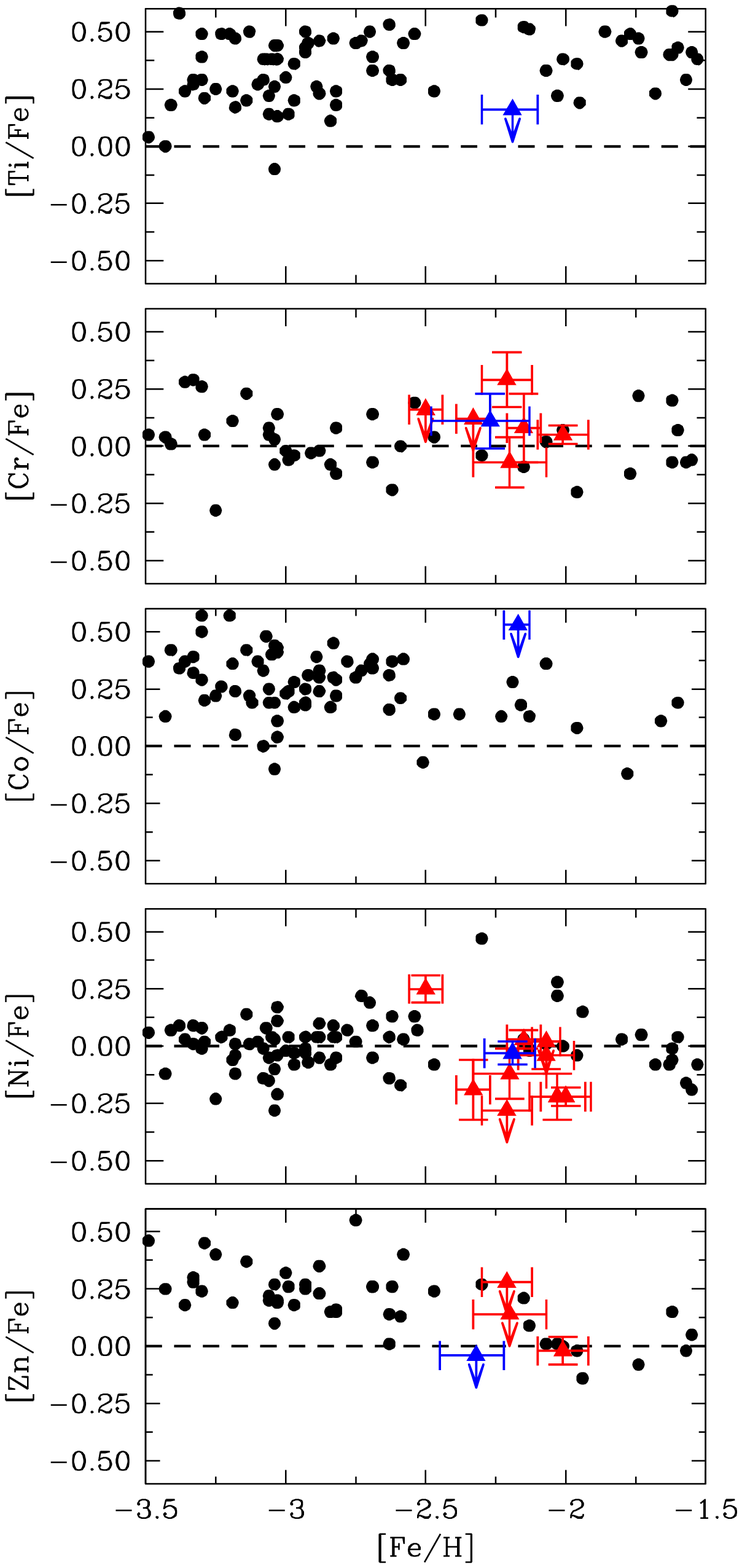}\\
  \caption{
Comparing Fe-peak element ratios in VMP DLAs (triangles)
and stars (filled circles;
\citealt{Gra03}, \citealt{Cay04}, \citealt{Lai08}, \citealt{Bon09},
and \citealt{BerPicGeh10}).
The red triangles are for
individual DLAs where the relevant absorption lines were
detected, or where an informative upper limit
could be obtained. The single blue triangle in each panel
corresponds to the stack of DLAs in which no absorption
by the element of interest 
was detected in individual systems.
  }
  \label{fig:compstars}
\end{figure}

To obtain independent confirmations of the stellar trends, 
we compare in Fig.~\ref{fig:compstars}
the Fe-peak ratios Ti,\,Cr,\,Ni,\,Zn/Fe 
measured in our sample of VMP DLAs
(filled triangles)
with the values found in the halo star
surveys (filled circles) by
\citet{Gra03}, \citet{Cay04}, \citet{Lai08}, \citet{Bon09}
and \citet{BerPicGeh10}.
In all cases except Cr/Fe, stellar abundances were
derived from 1-D LTE models. For the Cr/Fe ratio
we have instead used the one dimensional non-LTE (NLTE)
estimate by \citet{Bon09}. Where necessary,
the stellar measurements have been corrected
to the abundance scale adopted in this work
(see Section~\ref{sec:stacktech}
and Table~\ref{tab:solabu}).

Overlaid on the five panels 
of Fig.~\ref{fig:compstars}
are the DLA measurements and upper limits 
from Table~\ref{tab:summary}.
The red triangles are for DLAs with detections of
Fe-peak element lines, 
or where a useful upper limit could be derived.
The single blue triangle in each
panel shows the mean value of [X/Fe] for
the population of VMP DLAs without individual detections.
The error bars on the values of [X/Fe] 
include both random and systematic
uncertainties, and the ranges of [Fe/H] to which
the stacks of undetected lines refer were estimated
as explained in Section~\ref{sec:nodetect}.
We now comment separately on the
individual Fe-peak element ratios.

\subsection{Titanium}

There are two resonance lines of \TiII\
accessible in the rest-frame ultraviolet, at
$\lambda 1910.6123$ and $\lambda 1910.9538$.
The corresponding velocity separation,
$\Delta v = 53.7$\,km s$^{-1}$ is sufficient
to avoid contamination between the two spectral
regions over which we integrate, of width
$\Delta v = \pm 25$\,km s$^{-1}$ centred on each 
line. In any case, any contamination would
have the result of making the upper limit 
we deduce for the Ti/Fe abundance more conservative.
As can be seen from Fig.~\ref{fig:Nstack} (second row, left panel)
even by stacking 16 \TiII\ lines we do not achieve
a detection; the $3 \sigma$ upper limit we deduce,
[Ti/Fe]\,$\leq +0.16$ at [Fe/H]\,$\simeq -2.2 \pm 0.1$
(Table~\ref{tab:summary})
is plotted as a blue triangle in the top panel of
Fig.~\ref{fig:compstars}.

The derived limit on [Ti/Fe] is close to the lower
envelope of [Ti/Fe] values in stars of
comparable metallicity, derived from  the analysis of \TiII\
lines\footnote{We ignore stellar abundances
derived from \TiI, as per the recommendation by
\citet{Bre11}.}. This does not leave much room
for agreement between stars and DLAs, and the
lack of any discernible absorption beyond the $1\sigma$ region
near $0$ km s$^{-1}$ in the Ti/Fe panel of Fig.~\ref{fig:Nstack}
further suggests some
level of disagreement. Departures from
LTE may result in 
small corrections to the stellar Ti/Fe ratios 
measured from  \TiII\ lines \citep{Bre11}.
This may explain some of the mismatch between
stars and DLAs. An alternative, or additional,
possibility is that there may be 
a small residual degree of Ti depletion
onto dust grains in VMP DLAs. Although it is
usually assumed that DLAs with metallicities
$\lesssim1/100$ of solar exhibit negligible
depletion \citep{Pet97,Ake05}, 
Ti is one of the most refractory elements,
and is readily incorporated into dust grains
in the local interstellar medium
(see e.g. \citealt{SpiJen75}).
We thus caution that Ti \emph{may} suffer from a
mild depletion in DLAs when [Fe/H] $\simeq -2.0$.
However, given that the upper limit of Ti/Fe
derived here for the most metal-poor DLAs is consistent
with the values obtained from explosive nucleosynthesis
calculations of core-collapse supernovae
(see Section~\ref{sec:models} and Figure~\ref{fig:expenergy}),
with no appreciable $\alpha$-enhancement, we suggest
that Ti in the most metal-poor DLAs behaves like
an Fe-peak element rather than an $\alpha$-capture element.

\subsection{Chromium}
Chromium is accessible via the well known triplet 
\CrII\,$\lambda\lambda 2056, 2062, 2066$ \citep[e.g.][]{Pet90}.
In the present sample, we have four measures
of [Cr/Fe] and two upper limits in individual
DLAs, and a $5 \sigma$ detection in the stack
of DLAs without individual detections
(see Table~\ref{tab:summary}).
These seven determinations are all in 
good mutual agreement,
and they broadly agree with the stellar data,
as can be seen in the
second panel from the top in Fig.~\ref{fig:compstars}.
In the DLAs, [Cr/Fe] is approximately solar,
or possibly slightly supersolar by $\sim 0.1$\, dex,
down to [Fe/H]\,$\simeq -2.5$. 

In stars the Cr/Fe ratio is found to be roughly solar
down to [Fe/H] $\sim-2.2$. For [Fe/H] $\lesssim-2.2$,
the earliest measures of Cr/Fe (based on the \CrI\ lines)
suggested that this ratio gradually decreases to
[Cr/Fe]\,$\simeq-0.5$ when [Fe/H]\,$= -3.5$ \citep{Cay04}.
This decline may however be spurious,
reflecting an increasing correction with
decreasing metallicity for NLTE effects
to the \CrI\ lines used in the analysis.
On the basis of their NLTE calculations,
\citet{BreCes10} have argued that
the [Cr/Fe] ratio remains close to solar down to
the lowest metallicities, in good
agreement with the Cr/Fe ratio in metal-poor stars
measured from the \CrII\ lines
(\citealt{Lai08,Bon09}, as plotted in Fig.~\ref{fig:compstars}).
Unfortunately, our survey does not extend
to lower metallicities where the stellar data
may deviate from solar Cr/Fe.
In future, it should be possible
to verify the behaviour of the Cr/Fe ratio at the
lowest metallicities with observations of VMP DLAs
of lower metallicity than those considered here.
This is an important check to perform
because models of Population III nucleosynthesis
interpret  sub-solar Cr/Fe ratios as resulting from a deeper mass
cut (i.e. larger explosion energy; \citealt{UmeNom02}).
Such an explanation is in line with the observed rise
in both the Co/Fe and Zn/Fe ratios 
at metallicities [Fe/H]\,$\lesssim -2.5$ (see below).

\subsection{Cobalt}
Together with Ti, Co is the least common of the
Fe-peak elements considered here, with an abundance
of only $\sim 1/350$ of that of Fe in the Sun
(Table~\ref{tab:solabu}). Thus, it is not surprising
that we do not detect \CoII\ absorption in any
of the 25 VMP DLAs in the present sample.
This difficulty is offset to some degree by
the availability of many \CoII\ transitions
in the wavelength interval 1400--2000\,\AA\ 
which is well observed in DLAs at $z = 2$--3
(see Table~\ref{tab:atomic}). By stacking a total
of 85 such features we are able to set an instructive
$3\sigma$ upper limit of [Co/Fe] $\le+0.53$
for the VMP DLA population at a mean 
metallicity [Fe/H]\,$\simeq -2.2$ (Table~\ref{tab:summary}).

The earliest compilations of element ratios in the most
metal-deficient stars indicated that Co is  enhanced
relative to Fe when [Fe/H] $< -2.0$ \citep{RyaNorBes91,McW95},
if the Co abundance is measured from \CoI\ lines assuming LTE, 
as  is the case for the data in Fig.~\ref{fig:compstars}. 
Even though later studies
have confirmed the trend of increasing [Co/Fe] with decreasing
metallicity \citep{Cay04,Lai08,Bon09}, the true level of the enhancement
of Co relative to Fe is still a matter of debate.
The \CoI\ lines appear to suffer
from large (positive) corrections due to the neglect of modelling
their hyperfine structure (HFS) as well as departures
from LTE \citep{BerPicGeh10}. 
In some cases, the offset is calculated to be as large as $+0.6$ dex. 
Applying these corrections
would result in a \emph{substantial} enhancement of Co, 
to a level that may be inconsistent with the upper limit
to [Co/Fe] we have determined in the DLAs
(see middle panel of Fig.~\ref{fig:compstars}).
Furthermore, model calculations of Co production in
CCSNe have difficulty in reproducing 
such extreme enhancements of the Co/Fe ratio.
Some authors have interpreted the rise in [Co/Fe]
at low metallicities as evidence for
high energy CCSN enrichment \citep{UmeNom02}, but even then
there is still a shortfall in the model abundances compared 
to the stellar values corrected for NLTE and HFS effects.
DLAs can provide new insights into this problem, and will undoubtedly
play an important role in future to help settle the debate.

\subsection{Nickel}
Ni is the second most abundant Fe-peak
element in the Universe. This fact, in addition
to the multitude of relatively strong transitions
that are easily accessible (i.e. the lines are
usually free from both the \Lya\ forest and
telluric absorption), has resulted in Ni being
the most commonly observed Fe-peak element in
high redshift DLAs (besides Fe). For the 
sample of metal-poor DLAs considered here, we have
seven detections and two upper limits in individual
systems, and a total of 52 lines contributing to the 
stack of undetected lines. 

%%%%%%%%%%%%
% FIGURE 5 %
%%%%%%%%%%%%
\begin{figure*}
  \centering
  {\hspace{-0.25cm} \includegraphics[angle=0,width=80mm]{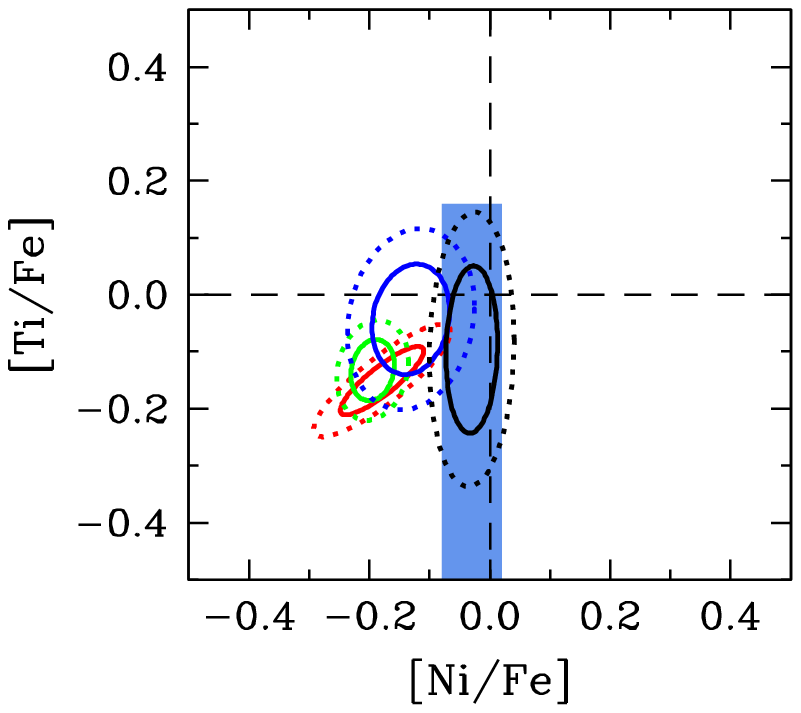}}
 {\hspace{0.25cm} \includegraphics[angle=0,width=80mm]{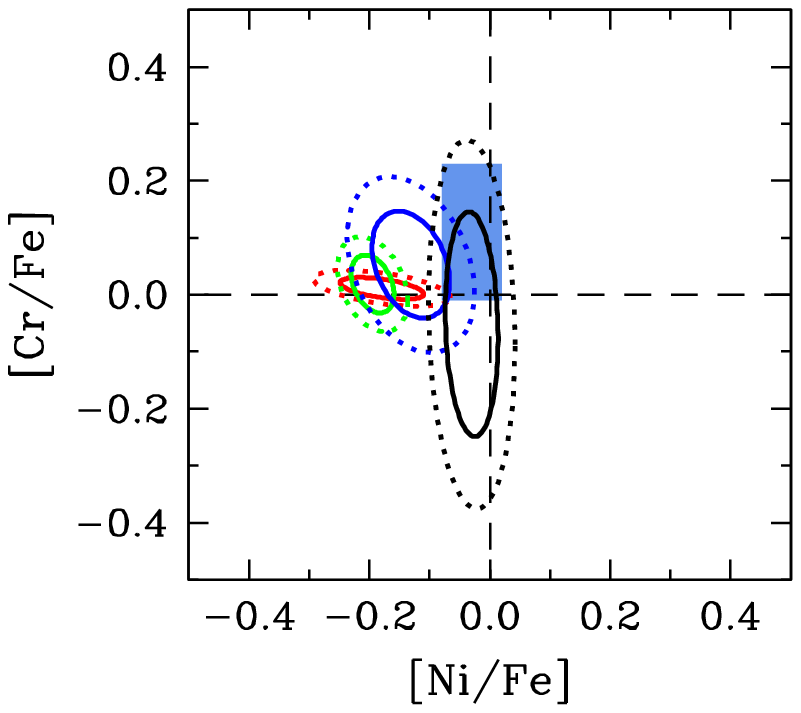}}\\
 {\vspace{0.5cm}}
 {\hspace{-0.25cm} \includegraphics[angle=0,width=80mm]{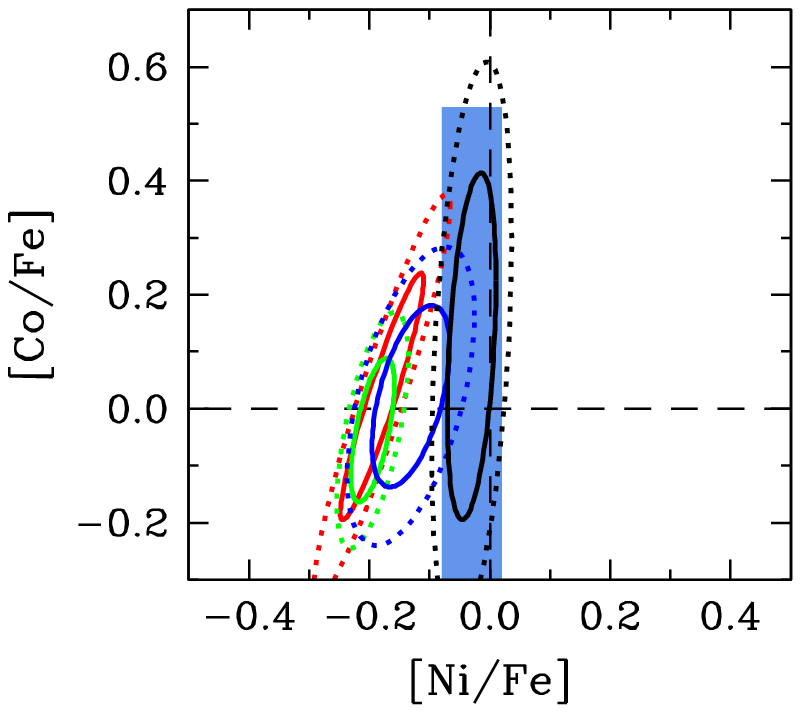}}
 {\hspace{0.25cm} \includegraphics[angle=0,width=80mm]{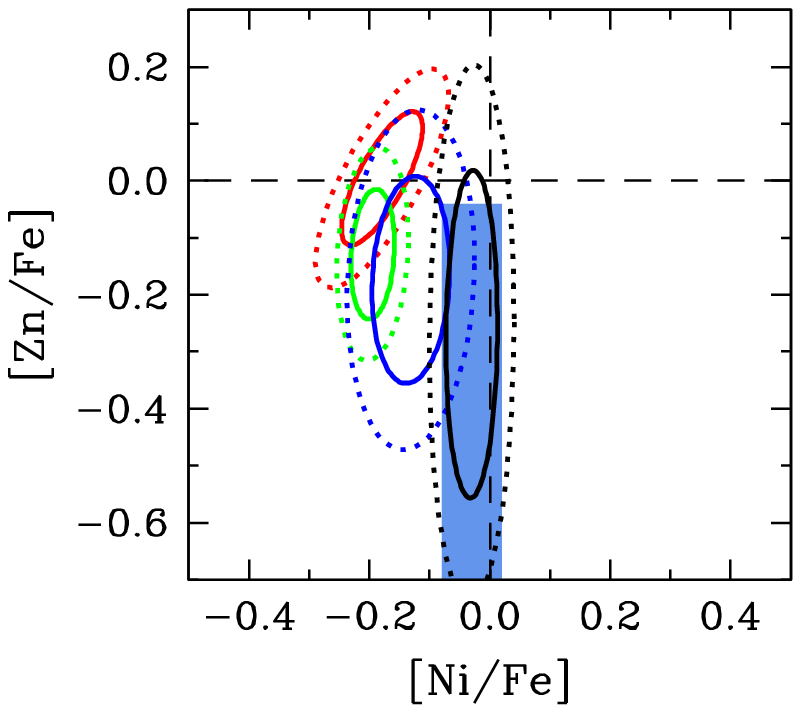}}\\
  \caption{
Comparison of the explosive nucleosynthesis calculations by
\citet{HegWoo10} with elements ratios measured in VMP DLAs.
In each two-ratio plane, the theoretical values 
are shown with ellipses enclosing 68\% (solid ellipses) 
and 95\% (dotted ellipses)
of the models considered in 
each band of supernova explosion energy.
Four bands are considered, colour-coded red, green, blue, and black.
Broadly speaking, red corresponds to hypernovae, green is for high
energy Type II supernovae, blue represents a typical Type II supernova,
and black corresponds to the faint supernovae 
(see text for further details and the
specific cuts we have employed). 
The blue shaded region in each panel
shows the values measured in the stack of VMP DLAs.
  }
  \label{fig:expenergy}
\end{figure*}

In Galactic  stars, the relative abundances of Ni and Fe
remain essentially constant over the entire
range of metallicity probed. Although there are
 notable exceptions to the rule, the typical stellar
[Ni/Fe] value is solar
([$\langle$Ni/Fe$\rangle$]$=-0.01$), in
very good agreement with the value
[Ni/Fe]\,$= -0.03 \pm 0.04 \pm 0.03$
we derive for VMP DLAs at a mean [Fe/H]\,$\simeq -2.2$
(see Fig.~\ref{fig:compstars}). It is interesting,
however, that most of the systems with
\emph{detected} \NiII\ absorption exhibit
slightly sub-solar ratios, 
except for one DLA with an
unusually high Ni/Fe ratio.
It appears that both stars and DLAs show
significant variation in their Ni/Fe ratios, and
DLAs preferentially have sub-solar values of
[Ni/Fe].

\subsection{Zinc}\label{sec:starsZn}
Zn is rarely detected in DLAs when the metallicity is less than
1/100 of solar; we have only one detected system in our sample 
(reported by \citealt{Mol00}) in a DLA with 
one of the largest recorded values of 
neutral hydrogen column density, 
$\log N$(\HI)/cm$^{-2} = 21.41$.
In addition,
there are a further two systems in the literature with useful upper
limits. From the stacked spectrum of 11 unblended \ZnII\
lines in DLAs  without individual 
detections, we deduce a $3\sigma$ upper
limit [Zn/Fe]\,$\le -0.04$ at a mean [Fe/H]\,$\simeq -2.3$.

While [Zn/Fe] is close to solar in stars with 
[Fe/H]\,$> -2.0$ \citep{Nis04,Nis07}, it rises to
supersolar values at lower metallicities (see bottom
panel of Fig.~\ref{fig:compstars}). This trend,
first noted by \citet{Pri00} and 
later confirmed by \citet{Cay04} with a larger sample of halo stars,
is unlikely to be due to the details of the stellar atmosphere modelling.
The currently favoured interpretation,  suggested by
\citet{UmeNom02}, is that it reflects nucleosynthesis by 
hypernovae --- supernovae with an order of magnitude
larger explosion energy than normal. The higher energy explosion
results in a stronger $\alpha$-rich freezeout, which in turn
favours the production of Zn.

Although the statistics for VMP DLAs are limited, our 
$3 \sigma$ upper limit [Zn/Fe]\,$\le -0.04$ for the 
population of DLAs with a mean [Fe/H]\,$\simeq -2.3$
(blue triangle in the bottom panel of 
Fig.~\ref{fig:compstars}) is not consistent
with an enhancement of Zn at these metallicities.
There is no hint of \ZnII\ absorption
in the stack shown in the bottom panel of Fig.~\ref{fig:Nstack},
suggesting that the true value of [Zn/Fe] in VMP DLAs
may well be sub-solar and thus different from
the values found in stars of comparable metallicity. 
Corrections due to dust depletion effects or
unseen ion stages, although expected to be small,
would both act to lower the Zn/Fe ratio compared to
the value listed in Table~\ref{tab:summary} and plotted in Fig.~\ref{fig:compstars}.
Given that Zn has the potential to be a powerful probe of the explosion
energy, it is now important to increase the sample of DLAs with
measurable [Zn/Fe] down to at least [Fe/H]\,$=-3.0$
for a more instructive comparison with the stellar data than
is possible at the moment.
A full accounting of the relevant non-LTE
effects is also required to firmly establish the elevated
Zn/Fe ratio that is seen in stars at the lowest metallicities.

\section{explosion energy of Type-II supernovae}
\label{sec:models}

It was shown in \citet{Coo11b} that the
chemical composition of the most metal-poor
DLAs is consistent with the hypothesis
that they were enriched by massive stars 
that ended their lives as core-collapse supernovae. 
Specifically, the enhanced [O/Fe] ratio that
is observed for all of the DLAs in our sample, implies
that the dominant sources of metals in these systems
are Type-II supernovae.
In this section, we will attempt
to estimate the typical explosion energy of these supernovae 
by comparing the abundances of  Fe-peak elements 
deduced above with the relative yields calculated 
with models of explosive nucleosynthesis in metal-free stars.
It has been known for over a decade that the
relative abundances of the Fe-peak elements at low
metallicity can provide clues regarding
the physics of a core-collapse
supernova explosion \citep{WooWea95, UmeNom02}.
Progress in this area is highly desirable, 
because we still lack a physical understanding
of the \emph{mechanism} that
drives the explosion, despite many recent efforts
(e.g. \citealt{Mez05,OttOCoDas11}).
In the current generation of massive star nucleosynthesis models
this shortcoming is circumvented by
parameterising the physics of
the explosion in order to calculate the resulting
element yields. The parameterisation
is typically achieved by employing a simple description
of the mixing that occurs between the stellar layers
(due to either Rayleigh-Taylor or rotationally-induced
mixing), and specifying a mass-coordinate (or mass-cut)
that defines the material that either escapes the
binding energy of the exploding star, or falls back onto
the newly formed compact object.

Perhaps the most detailed CCSN explosive nucleosynthesis
calculations are those by \citet{HegWoo10}. These
authors have investigated the explosive nucleosynthesis in massive
stars ($M = 10$--$100$ M$_{\odot}$) with a fine mass-resolution
($\gtrsim0.1$ M$_{\odot}$); the entire suite consists of 120 stellar models.
Each star undergoes a supernova explosion, which is simulated as a moving
piston that deposits momentum at a specified mass-coordinate. We have adopted
the standard case which sets the piston to be at the base of the oxygen burning
shell, corresponding to an entropy per baryon of $\simeq4\,k_{\rm B}$. Each exploding star
is simulated with 10 different values of the final kinetic energy at infinity, taken to
be in the range ($0.3-10$) $\times10^{51}$ erg. A simple prescription of the mixing
that occurs between the stellar layers during the explosion is also implemented.
This is achieved by running a boxcar filter with a variety of  widths (14 in total),
ranging from no mixing at all to the maximum filter width of 25\% of the He core size.
The entire suite of models consists of a total of 
16\,800 combinations of these parameters.

For a given final kinetic energy at infinity and amount of mixing, we weight
these chemical yields by a power-law stellar initial mass function (IMF) with an
exponent that can take values in the range 0.35--2.35, where a Salpeter-like IMF
corresponds to 1.35. Our conclusions on the explosion energy
therefore apply to the mass range of stars
with the dominant Fe-peak element yields.
We subdivide the weighted
yields into four arbitrary groups depending on the final kinetic energy at infinity:
(1) the hypernovae, with energies in the range (5.0--$10.0)\times 10^{51}$\,erg;
(2) the high-energy supernovae, with energies in the range (1.8--$3.0) \times 10^{51}$\,erg;
(3) the typical supernovae, with energies in the range (0.9--1.5)$\times 10^{51}$\,erg; and
(4) the faint supernovae, with energies in the range (0.3--$0.6)\times 10^{51}$\,erg.
In Fig.~\ref{fig:expenergy}, these four 
groups are colour-coded in red, green, blue, and black respectively.
Given the large number of models
considered, in Fig.~\ref{fig:expenergy}
we represent each of these four explosion energy ranges with two
ellipses that enclose, respectively, 68\% (solid lines) and 95\% (dotted lines) 
of the models in that range. In this way, we can learn how likely a certain
explosion energy is, given two measured Fe-peak element ratios.
For all panels in the figure, we also show the ranges of  Fe-peak
element ratios determined from the DLA stacks
(displayed as shaded boxes).

In all of the panels for Fig.~\ref{fig:expenergy},
it is evident that [Ni/Fe] has the strongest discriminatory power on
separating the low and high energy supernovae.
The abundances of Ti, Cr, and Co relative to iron,
on the other hand, are similar for all four
energy groups. Nevertheless, it is
reassuring that the model calculations and
VMP DLA measures (or upper limits)
of the Ti/Fe, Cr/Fe and Co/Fe
ratios are in good agreement with each other.

%%%%%%%%%%%%
% FIGURE 6 %
%%%%%%%%%%%%
\begin{figure}
  \centering
  \includegraphics[angle=0,width=80mm]{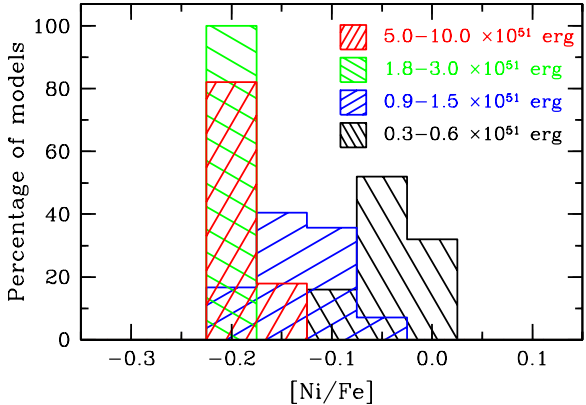}\\
  \caption{
The distribution of [Ni/Fe] values in the IMF-weighted
yields for the \citet{HegWoo10} models.
The four sets of models are colour-coded according to the 
kinetic energy released by the supernovae.
  }
  \label{fig:NiFedist}
\end{figure}

As can be seen from the bottom right panel of
Fig.~\ref{fig:expenergy}, it is the combination
of the Zn/Fe and Ni/Fe ratios that provides the
most discriminatory power on the explosion energy.
The relationship between Zn/Fe and explosion energy
has been previously noted by
\citet{UmeNom02}. These authors invoked the increased importance
of hypernovae with decreasing metallicity to explain the gradual rise to
[Zn/Fe]\,$\simeq  + 0.7$ at [Fe/H]\,$\sim- 4.0$ discovered in
the most metal-poor halo stars \citep{Pri00,Cay04}. 
However, even the most energetic
supernovae in the \citet{HegWoo10} models underproduce Zn 
compared to the enhanced values
reported by  \citet{Cay04} and \citet{Lai08}.
To account for this deficiency, \citet{HegWoo10} caution that
Zn can also be produced by other (unaccounted for) sources, such
as the neutrino wind from a newly formed neutron star (i.e. the `hot bubble'
component; for a relevant discussion, see \citealt{Pru05}).
\citet{IzuUme10}, however, point out that such
a scenario requires a contrived degree of fine-tuning for the
explosion parameters and favour hypernovae
as the more natural explanation for
the elevated Zn/Fe ratios.

Thus, the jury is still out regarding the origin of the 
high Zn/Fe ratios measured in the most-metal poor stars.
As has been shown to be the case for other elements
\citep[e.g.][]{Coo11b},  complementary observations
in VMP DLAs  have the potential of resolving the impasse.
Unfortunately, the \ZnII\ lines in VMP DLAs
are very weak, and their detection with current instrumentation
will require the rare combination of bright background QSOs
and unusually high
column densities of neutral hydrogen. We estimate that
DLAs with \NHI\,$> 10^{21}$\,cm$^{-2}$
are required to have a chance of detecting the 
\ZnII\ lines in  the metallicity regime [Fe/H]\,$<-2.5$;
such DLAs do exist, but all known cases are located 
in front of QSOs too faint for their spectra 
to be recorded at the required
S/N and resolution with current instrumentation.
However, measuring the Zn/Fe ratio in the most metal-poor 
DLAs is a goal that can be readily achieved
with the forthcoming generation of 30-m class telescopes.
This is a very exciting prospect. Only by measuring the
relative abundance of the Fe-peak elements in DLAs with [Fe/H]\,$<-3.0$,
can we confirm the trend of [Zn/Fe]  at the lowest metallicities,
and thereby pin down the typical energy released by
the CCSNe of the first stars.

Although the Zn/Fe ratio provides the strongest
separation between hypernovae and other types of SNe,
it cannot discriminate the typical SNe from faint SNe.
The clear, but weak dependence of [Ni/Fe] on the explosion
energy, which has not previously been noted, provides the
required separation. To illustrate this point better we show
in Fig.~\ref{fig:NiFedist} the full distribution of [Ni/Fe] values derived 
from the Heger \&\ Woosley (2010) models with the same colour-coding
as in Fig.~\ref{fig:expenergy}. Thus, the combination of these three
elements, Fe, Ni and Zn, offers the prospect of pinning down the
explosion energy of the stars that produced most of the Fe-peak
elements in the most metal-poor DLAs.

The primary motivation of this work was to 
estimate the energy released by the stars
that enriched the most metal-poor DLAs. 
The results presented here are only a first step in this
direction and have mostly highlighted the need for 
further observations of DLAs at even lower metallicities than
the regime probed so far.
Based on the low value of  [Zn/Fe]
and near-solar values of [Ni/Fe] we have found,
we tentatively conclude that VMP DLAs were probably 
enriched by massive stars that released
$\lesssim1.2\times10^{51}$ erg of energy, 
However, we caution that this conclusion is based on the assumption
that these absorbers were enriched only by Population III stars
(and that the nucleosynthesis by such stars is accurately reproduced
by current models).
It  remains to be tested how the relative Ni, Zn and Fe yields
depend on the metallicity of the progenitor star, as well as  
the explosion energy. This is a goal for future, high mass-resolution, 
metallicity-dependent stellar yield calculations.

\section{Summary and Conclusions}
\label{sec:conc}

We have presented the first dedicated survey of 
Fe-peak elements in very metal-poor DLAs 
at redshifts $z \sim 2$--3,
focussing in particular on the ratios of Ti, Cr, Co, Ni, Zn 
to Fe.
Based on the analysis of the relative abundances of these
elements in a sample of 
25 DLAs with [Fe/H]\,$\le 2.0$,
we draw the following conclusions:

\noindent ~~(i) In nine of the 25 DLAs,  we could measure
element ratios, or useful upper limits, directly from 
detected absorption lines. In the majority of DLAs, however,
the lines are too weak to be detected in available spectra.
To deal with such cases, 
we have devised a form of spectral stacking that
allows us to measure the Fe-peak element ratios
typical of the VMP DLA population as a whole,
at mean metallicities [Fe/H]\,$ \simeq -2.2$.

%\smallskip

\noindent ~~(ii) Comparing the Fe-peak element ratios
deduced in VMP DLAs with the values measured
in metal-poor stars of the Galactic halo, we found
fair agreement for [Cr/Fe] and [Ni/Fe].  
 Furthermore, the upper
limit on [Co/Fe] in VMP DLAs  
is consistent with the stellar values,
provided  the non-LTE corrections to the latter
are not too severe. On the other hand,
the  upper limits we place on the [Ti/Fe] and [Zn/Fe] values
in VMP DLAs are barely consistent with the stellar measurements; 
the DLA upper limits lie near the lower envelope of
these ratios in halo stars. Whether there is indeed a 
discrepancy or not can only be confirmed by future
observations of  DLAs
with [Fe/H]\,$< -2.5$ , this being the metallicity regime 
where the stellar trends exhibit marked
deviations from the solar ratios.

%\smallskip

\noindent ~~(iii) By comparing the Fe-peak element ratios 
in the DLAs to the values computed with the most detailed 
current models of nucleosynthesis by metal-free
stars, we tentatively concluded that the
DLAs in our sample were probably enriched by massive stars 
that released $\lesssim1.2\times10^{51}$\,erg 
when they exploded as core-collapse supernovae.
An extension of models such as those by \citet{HegWoo10}
to Population II stars is required 
to test if the Fe-peak element yields, particularly
those of Ni, Zn and Fe, depend on metallicity
as well as the stellar mass and explosion parameters.

%\smallskip

The present belief is that the most metal-poor Galactic stars contain
the metals synthesised by an earlier generation of massive stars 
that had ended their lives as hypernovae;
however, we have found little evidence so far for this
hypothesis in the chemical composition of the most metal-poor DLAs.
Future observations with the forthcoming generation of
30-m class telescopes, equipped with echelle spectrographs, will
allow us to confidently measure the relative abundances of
Fe-peak elements in individual  DLAs with [Fe/H]\,$<1/1000$ of solar.
We suspect that only then will we be able to reliably determine the
energy released by the first stars.

\section*{Acknowledgements}
We are grateful to 
the staff astronomers at the VLT and Keck  
Observatories for their assistance 
with the observations,
and to the telescope time assignment
committees for their support of the VMP DLA survey.
We thank the referee, Paolo Molaro, for a prompt
and helpful report that improved the paper, and
Piercarlo Bonifacio for communicating his measurements
of Cr/Fe in very metal-poor stars.
It is also a pleasure to acknowledge valuable
discussions with Poul Nissen and Stan Woosley,
regarding the stellar abundance measurements
and nucleosynthesis calculations.
We thank the Hawaiian
people for the opportunity to observe from Mauna Kea;
without their hospitality, this work would not have been possible.
RC acknowledges the support of a Morrison Fellowship provided
by the University of California, Santa Cruz.
RAJ is supported by an NSF Astronomy and Astrophysics
Postdoctoral Fellowship under award AST-1102683.
MTM thanks the Australian Research Council
for a QEII Research Fellowship (DP0877998).

%\bsp

%\begin{appendix}

%\end{appendix}

\label{lastpage}

\end{document}